\documentclass[letterpaper,twocolumn,10pt]{article}

\usepackage{usenix}
\usepackage[pdftex]{graphicx}
\graphicspath{{../images/}}
\DeclareGraphicsExtensions{.pdf,.jpeg,.png}
\usepackage{array}
\hypersetup{
    colorlinks=true,
    linkcolor=violet,
    urlcolor=cyan,
    citecolor=black,
    pdfpagemode=FullScreen,
}
\usepackage{amsmath}
\usepackage{amsfonts}
\usepackage{tikz}
\usepackage{etoolbox}
\usepackage[ruled,vlined]{algorithm2e}
\usepackage{booktabs}
\usepackage{caption}
\usepackage{tabularx}
\usepackage{makecell}
\usepackage{pgfplots}
\usepackage{pgfplotstable}
\usepackage{filecontents}
\usepackage{subcaption}

\usepackage[capitalize,nameinlink]{cleveref}

\usepgfplotslibrary{groupplots}
\usetikzlibrary{pgfplots.groupplots}
\usetikzlibrary{matrix}
\usepackage{placeins}
\usepackage{enumitem}

\newcommand{\att}{%
    \textit{ASTRA}%
}

\renewcommand{\eqref}[1]{\mbox{Eq.~\ref{#1}}}

\begin{document}

\title{\Large \bf \textit{May I have your Attention?} Breaking Fine-Tuning based Prompt Injection Defenses using Architecture-Aware Attacks}

\author{
{\rm Nishit V. Pandya}\\
UC San Diego
\and
{\rm Andrey Labunets}\\
UC San Diego
\and
{\rm Sicun Gao}\\
UC San Diego
\and
{\rm Earlence Fernandes}\\
UC San Diego
}

\maketitle

\pgfplotsset{compat=1.18}

\begin{abstract}
A popular class of defenses against prompt injection attacks on large language models (LLMs) relies on fine-tuning to separate instructions and data, so that the LLM does not follow instructions that might be present with data. 
%There are several academic systems and production-level implementations of this idea. 
We evaluate the robustness of this approach
%class of prompt injection defenses 
in the whitebox setting by constructing strong optimization-based attacks, and show that the defenses do not provide the claimed security properties. Specifically, we construct a novel attention-based attack algorithm for textual LLMs and apply it to three recent whitebox defenses SecAlign (CCS 2025), SecAlign++, and StruQ (USENIX Security 2025), showing attacks with success rates of up to \textbf{85-95\%} on unseen prompts with modest increase in attacker budget in terms of tokens. Our findings make fundamental progress towards understanding the robustness of prompt injection defenses in the whitebox setting. We release our code and attacks at \href{https://github.com/nishitvp/better_opts_attacks}{this link}.
\end{abstract}

\section{Introduction}
\label{sec:intro}

%LLMs becoming important. Prompt injections major threat. Give real world examples of prompt injection attacks. Mention how fundamentally prompt injections happen because can't separate instructions from data. To combat prompt injection attacks, people proposed fine-tuning based defences to train models to separate instructions from data. Three notable examples of such models include Secalign, Struq, Instruction Hierarchy. Notably, secalign and struq show high resistance even against best currently-known strong optimization-based attacks.
%In this paper, we present new attack that can break the white box defences with high success rate.
%Our attack relies on exploiting the specific architectural props, ignored by previous generic attacks.

Prompt injection attacks are one of the most pressing security issues in Large Language Models (LLMs). They are akin to stack smashing and confuse the LLM into following instructions that appear in data ~\cite{zverev2025can}. With the emergence of so-called agentic systems, in which LLMs can use tools to automate a variety of tasks~\cite{yao2023react, Schick2023ToolformerLM}, it is critical to defend against prompt injection attacks. 

%To address this problem on a fundamental level, many 
Recent work has proposed fine-tuning based defenses that train models to learn special tokens that can delimit instructions and data. Notable examples include Meta's SecAlign~\cite{chen2025secaligndefendingpromptinjection} and SecAlign++ ~\cite{chen2025metasecalignsecurefoundation}, OpenAI's instruction hierarchy~\cite{wallace2024instructionhierarchytrainingllms},  and an academic system called StruQ~\cite{chen2024struqdefendingpromptinjection}. Our work contributes an experimental security analysis of this class of prompt injection defenses in the whitebox setting. Particularly, we focus on SecAlign (CCS 2025), SecAlign++ and StruQ (USENIX Security 2025), which have showed high resistance to prior state-of-the-art optimization-based attacks such as Greedy Coordinate Gradient (GCG) and AdvPrompter \cite{zou2023universaltransferableadversarialattacks,paulus2025advprompterfastadaptiveadversarial}.
%\footnote{Note: We do not attack OpenAI instruction hierarchy because it is a closed implementation. However, this defense operates on the same principles as StruQ and SecAlign.}

As our first contribution, we experimentally establish that GCG-style attacks are not strong enough to fully vet the security of such defenses, showing that they do not perform much better than a simpler greedy search procedure.

% this thing is redundant: we already mention results later in intro.
%and apply it to attack SecAlign and StruQ. We achieve attack success rates of upto $60-80\%$ on unseen prompts on the evaluation sets proposed by the authors of these systems, showing that the defenses do not provide the claimed security benefits under an attacker who uses architectural information even under a weak threat model. \att~uses a modest increase in budget as measured by the number of tokens injected into the prompt. 

Our second contribution is a novel architecture-aware attack algorithm, \att\ (short for \textit{Adversarial Subversion through Targeted Redirection of Attention}),\footnote{``Astra'' also means ``weapon'' in several Indian languages.} that targets the attention matrices within transformer-based LLMs. Using \att, we conduct a stronger and more realistic evaluation of fine-tuning based defenses. Although StruQ and SecAlign do evaluate their security against an adaptive adversary, their evaluation is limited by a lack of strong optimization-based attack algorithms for LLMs. This is a critical learning from the first wave of adversarial machine learning research -- defenses need to be evaluated against strong, adaptive adversaries which take into account the defense's principle of operation~\cite{athalye2018obfuscatedgradientsfalsesense,carlini2019evaluatingadversarialrobustness,tramer2020adaptiveattacksadversarialexample}. \att\ allows for precisely this kind of an evaluation.

\att~enables a stronger adversary than GCG because (1) it uses the whitebox setting better by exploiting the architecture of the model to craft attacks (2) it can create attacks that are universal across the contents of the context window -- this is especially relevant to agentic use cases of LLMs and goes beyond the original evaluations of StruQ and SecAlign (3) it allows us to effectively evaluate the defense by scaling the attacker budget (measured in number of injected tokens).

\att~finds adversarial input tokens that try to cause the LLM to focus its attention -- as measured by internal attention matrices -- on the attacker-supplied instructions (i.e., the injected prompt). Intuitively, if the LLM's context window only had the attacker's instructions and the attacker's instructions didn't violate a content safety policy, then the LLM would follow them because there are no other instructions in the context. Following this heuristic, our attack generates sequences of tokens that try to make the LLM ignore all other tokens in the context window, except the attacker-supplied ones. By contrast, the Greedy Coordinate Gradient (GCG) algorithm tries to find inputs that cause the LLM to generate attacker-desired output tokens, without taking the LLM's architectural structure into consideration. That is, the attacker's loss function is expressed in terms of output token probabilities. In practice, this represents a very difficult optimization objective, as evidenced by GCG's general inability to convincingly break defenses like SecAlign. On the other hand, \att~gains its power from architectural awareness --- by expressing the attacker's loss function in terms of attention, it is able to compute better attacks than GCG alone.

% \al{I am actually not entirely sure we need to determine the attention heads importance - by default, we can just act on all of them?}
% \np{It is not essential strictly, in the sense that some attacks do work without doing all this, but it does help to do it this way, at least as measured by the loss curves}

% \al{Even in the steering based attacks, we act on all attention heads, it is that the effect is necessary for only a few of those attention heads}
% \np{Even for steering, not all layers need to be steered. There too, there are some which are most essential. Also, steering based things act at the representation level, not at the attention head level}

% \al{We can alternatively say that existing architecture-aware attacks like steering-based don't work for our problem, but in fact we didn't try that, right?}
% \np{Shouldn't mention steering-based IMO.}

Expressing a prompt injection loss function in terms of the internal attention matrices of the model requires addressing the challenge of determining which attention heads (out of thousands in modern LLMs) are most important for the specific task that is given to the model. Attention heads are not independent, but rather, a change in an earlier head results in downstream changes. In general, solving the problem of determining which attention heads are important is a hard problem. It requires extensive manual intervention and/or multiple expensive ablations that involve turning-off attention heads or patching activations and empirically observing their effects on the task in question ~\cite{burns2023discovering,olsson2022incontextlearninginductionheads,zhang2024bestpracticesactivationpatching}. Our key insight is that in a prompt injection, the attacker wants specific output tokens to appear, and therefore, we do not need to explicitly find out which attention heads are critical for different types of fuzzily defined tasks. Rather, since our goal is to find out-of-distribution inputs, we need to determine the important attention heads for which small changes in the attention matrices cause large changes in the probability of the desired output tokens. We quantify the importance or weight assigned to each attention head using a novel sensitivity metric that can be computed using gradient information, thus bypassing the requirement for expensive ablations or manual interventions. 

A second challenge is evaluating the fine-tuning based defenses in a more realistic setting where the attacker does not have knowledge of the contents of the LLM's context window. This requires that the attacks produced by \att~be universal over unseen contexts. Although the original StruQ and SecAlign evaluations do not consider this scenario, we believe it is an important setting because it represents a more realistic attacker threat model. In general, an attacker will not have complete knowledge of the context window. For example, the context window might contain additional instructions, private user data, tool-calling demonstrations, etc. Furthermore, this setting is the norm, rather than the exception, in agentic use cases. To address this challenge, we create \att++, a stronger variant of \att\ that addresses the above challenge by dynamically adapting the weights assigned to each attention head during the optimization.

We evaluate \att~on SecAlign and StruQ. Using the same evaluation methodology as outlined in these papers, we demonstrate an attack success rate of \textbf{82.5\%} and \textbf{72.5\%} for the SecAlign defended versions of Mistral and Llama-3 respectively (obtained from the authors of these two systems). Our attacks require a modest increase in token budget --- i.e., the original StruQ and SecAlign papers evaluate the setting where the attacker is given $20$ tokens that are suffixed to the attacker's manually-written malicious instructions. We achieve high success rates by scaling the attacker's budget up to $35$ tokens that can be prefixed or suffixed (or both) to the injected instruction.

Going beyond the original settings of SecAlign and StruQ, we also evaluate \att++ in a more realistic setting where the adversary doesn't have knowledge of the surrounding context. With this constraint, the attacks generated by \att++ achieve a success rate of \textbf{96\%} and \textbf{87\%} on unseen contexts against the SecAlign and SecAlign++ defended versions of Llama respectively. Like before, we achieve high success rates by scaling the budget up to $35$ tokens.

StruQ and SecAlign only evaluate against an attacker limited to a 20-token budget. Unlike the computer vision space where L-norms represented attacker threat models (because they captured image distortion) and had natural limits, the LLM space has no such inherent restriction. Context windows of modern LLMs are very large (often up to tens of thousands of tokens) and the trend is to make them even larger. Furthermore, when an LLM is used as part of a larger agentic system, the attacker's instructions can arrive from any source, such as an email or a webpage which can easily span several hundreds of tokens, and thus, a larger attack string is likely to go un-noticed. On a fundamental level, the attacker is not restricted by the number of tokens they can inject.\\

\noindent\textbf{Contributions.}
\begin{itemize}[itemsep=0pt, parsep=0pt, topsep=0pt, partopsep=0pt]
    \item We empirically study the security of the class of fine-tuning based prompt injection defenses. We show that existing white-box optimization-based attacks fail to break these defenses because the gradient of a loss function that measures the output probabilities of target tokens does not convey sufficient information for optimization.

    \item We introduce \att\ and a closely related variant \att++, two novel optimization-based attacks that manipulate attention matrices to achieve prompt injection attacks. We show success rates of up to \textbf{85-95\%} on state-of-the-art SecAlign++, SecAlign and StruQ. The computed attacks are independent of prior tokens in the context window (i.e., our attack provides context window universality). 
    
\end{itemize}

\section{Background}
\label{sec:background}

\noindent \textbf{Transformer-based LLMs.} An LLM can be modelled as a function that takes in a sequence of tokens $(x_1, x_2, \dots x_n) \in V^{*}$ --- obtained by encoding an input string --- where each token is from a fixed vocabulary $V$, and outputs a probability distribution over the next token $P(\cdot \vert x_1, x_2, \dots x_n)$. The next token $x_{n+1}$ is generated by sampling from the distribution $P(\cdot \vert x_1, x_2, \dots x_n)$. The LLM then continues autoregressively, \textit{i.e.} it takes in $(x_1, x_2, \dots x_n, x_{n+1})$ as an input and generates $x_{n+2}$. This process continues until some stopping criterion is reached. The generated tokens are then decoded to get the final output string.

% We use the notation $LLM_{\Theta}(x_1, x_2, \dots x_n)$ to refer to the actual text generated by the LLM on the input sequence of tokens $(x_1, x_2, \dots x_n)$.

\noindent \textbf{Indirect Prompt Injection Attacks.} Indirect Prompt Injections are inference-time attacks that subvert the expected output of an LLM in order to cause the LLM to output text of the attacker's choice. Prompt injections aim to override the developer or user intent and instead cause the LLM to generate specific, malicious strings. Such attacks occur when an adversary controls a part of the input to an LLM. In real-world applications, an attacker can achieve such control by inserting malicious strings into third-party content that is placed into the context window of an existing conversation.

Formally, a prompt injection happens when a malicious string crafted by the adversary (say $x_{Adv}$) is combined with the trusted instruction of the user (say $x_{Trusted}$) and is processed by the LLM. In practice, strings from untrusted contexts often undergo special preprocessing, such as filtering, before they are combined with the trusted strings. They can also be delimited by other special tokens to mark separation between the trusted and untrusted contexts. The goal of the adversary is to craft an $x_{Adv}$ which when combined with $x_{Trusted}$ causes the LLM to output a string $y_{Target}$ of the attacker's choice, \textit{i.e.}
\begin{equation}
    LLM(\mathrm{Combine}(x_{Trusted}, x_{Adv})) \approx y_{Target}
    \label{eq:adv_goal_pi}
\end{equation}

Prompt injection attacks are of two types - handcrafted and automated. Handcrafted prompt injections exploit linguistic quirks, role-playing and other similar tricks. Automated prompt injections rely on an algorithm to generate adversarial inputs designed to mislead an LLM into generating attacker-chosen target strings. Most automated algorithms work by surrounding the attacker's malicious instruction (which we refer to as the ``payload'') with adversarial tokens~\cite{pasquini2024neuralexeclearningand,liu2024automaticuniversalpromptinjection,labunets2025fun}.

\noindent \textbf{Defenses against Indirect Prompt Injection.} The primary goal of a defense is to make a system secure against indirect prompt injection attacks. One major class of these defenses, detection-based ~\cite{liu2024formalizingbenchmarkingpromptinjection}, work by predicting if an input contains a prompt injection. The system may then reject a query if a prompt injection is detected. Examples of such defenses include Prompt Guard ~\cite{meta2024promptguard}, a known-answer detector and variants, like DataSentinel ~\cite{liu2025datasentinelgametheoreticdetectionprompt}.

Another major class of defenses addresses this problem by modifying the model directly so that it is more likely to correctly follow the original prompt instead of an injected instruction. These methods, which we refer to as \textit{fine-tuning based} defenses fine-tune the model itself to recognize the prompt structure and to separate instructions from data
Examples of such defenses include the open-source systems StruQ ~\cite{chen2024struqdefendingpromptinjection}, SecAlign ~\cite{chen2025secaligndefendingpromptinjection}, and Meta's SecAlign++ ~\cite{chen2025metasecalignsecurefoundation} and closed-source implementations such as OpenAI's Instruction Hierarchy ~\cite{wallace2024instructionhierarchytrainingllms}.
We describe a few examples of this class of defenses in more detail in section \cref{sec:secalign_struq}.

% more detection-based defenses: https://arxiv.org/abs/2310.12815
% more defencses: https://arxiv.org/pdf/2503.18813
% more defenses: https://arxiv.org/pdf/2504.11358

% Few different classes of defences proposed - 
% 1. Detection based defences
% 2. Training the model to separate instructions and data.
% a. Struq
% b. Secalign
% c. Instruction Hierarchy
% 3. Systems level defences

% \np{Is this needed?}
% Like for every other defense in the adversarial machine learning literature, evaluation of defenses against indirect prompt injections should be done against strong adaptive attacks ~\cite{carlini2019evaluatingadversarialrobustness, tramer2020adaptiveattacksadversarialexample, shi2025lessons}.
% In contrast with static evaluation against existing state-of-the-art attack strategies, evaluation against adaptive attacks adjusts the best known algorithms, their combination, and their parameters to target specifically the defense in question under reasonsable assumptions.
% Adaptive attacks emulate a stronger adversary and provide a more truthful lower bound estimate of security level under worst-case conditions.
% It is very common for an adaptive attack evaluation to identify edge cases, overlooked at the initial evaluation ~\cite{athalye2018obfuscatedgradientsfalsesense}.

\section{Threat Model}
\label{sec:threat_model}

Like prior work on indirect prompt injections, we assume that a (victim) user is using an LLM (target model) to interact with third-party content.

An attacker is an untrusted third-party who adversarially controls a part of the input to a target model and intends to steer the model away from following original, trusted instructions given by the user to a new set of instructions of the attacker's choice.
The delivery of adversarial instructions into a user's context window can be carried out via several different methods such as poisoning a webpage to be summarized (as instructed by the user's prompt to an AI assistant or as an operation of some agentic AI system), by poisoned documents, multimodal content, or code being processed ~\cite{wunderwuzzi2024zombaisc2claude,fu2023misusingtoolslargelanguage,greshake2023youve,emailAgentHijacking,bhatt2023purplellamacybersecevalsecure}.
For the purposes of this paper, we treat the real-world delivery mechanism as an orthogonal problem to our work. Nonetheless, we note that the variety of delivery mechanisms and the increasing capabilities of LLMs have important consequences for the broader, general threat model of a prompt injection attack. That is, with the trend of increasing context windows in LLMs and improving capabilities such as multiple language understanding, multi-modality, and tool calling abilities, the attacker is not, in general, restricted by constraints in number of tokens or the size of the attack itself.

We assume a whitebox, adaptive adversary, whose goal is to create prompt injection attacks with high likelihood of success. That is, we assume an adversary who has complete access to the weights of the target model and has all the details of the surrounding components that are orchestrating the LLM-user interaction. Examples of such components can be filtering or sanitization components that can filter text out of third-party content.

In the context of a prompt injection, there are two distinct assumptions one can make about the knowledge an adversary has about a user's conversation with an LLM.

\noindent \textbf{Weak Knowledge Assumption} - The adversary has \textit{no information} about the conversation history already present in the context window before the third-party data (controlled by the adversary) is injected. That is, the adversary knows nothing about $x_{Trusted}$ (\cref{eq:adv_goal_pi}). This assumption reflects a general prompt injection scenario for chat-based and agentic systems.

\noindent \textbf{Strong Knowledge Assumption} - The adversary has \textit{full information} of the exact conversation history before the third party data (controlled by the adversary) was placed into the context window - \textit{i.e.} the attacker knows $x_{Trusted}$. While unrealistic in the general case, there are instances where this strong assumption is valid. For example, if a user is using a task-specific bot with a fixed system prompt, such as the bots used in GPT-Store ~\cite{OpenAI2024GPTStore}, then an adversary can be reasonably expected to know the exact contents of the context window. Furthermore, this is the setting that StruQ and SecAlign used to evaluate the strength of their defenses.

% While SecAlign and StruQ evaluate their defenses against a strong knowledge assumption, we provide versions of \att~ under both the strong and weak knowledge assumptions.
\section{Fine-tuning based defenses}
\label{sec:secalign_struq}

In this section, we provide detailed background on three different fine-tuning based prompt-injection defenses - StruQ ~\cite{chen2024struqdefendingpromptinjection}, SecAlign ~\cite{chen2025secaligndefendingpromptinjection} and SecAlign++ ~\cite{chen2025metasecalignsecurefoundation}. Then, using SecAlign as a case study, we perform an analysis to explain why existing SOTA algorithms like GCG don't break it convincingly.

\subsection{Summary of some existing defenses}

\noindent \textbf{StruQ} starts with a base model and uses a version of adversarial training adapted for instruction tuning to obtain the final defended model. StruQ reserves special tokens which it uses to delimit data from instructions. The model is trained to recognize these delimiters and ignore the instructions present in the delimited part. The model is combined with a preprocessing filter that sanitizes input text to remove any occurences of these delimiters. The StruQ models resist hand-crafted prompt injection techniques with success rates close to $0\%$. Nonetheless, they are vulnerable to adversarial optimization methods like GCG, with attack success rates of approximately $60\%$ of the cases when using GCG.

\noindent\textbf{SecAlign} adopts a similar approach to StruQ. It uses special delimiters and preprocessing filters that removes special tokens from the untrusted data ingested by the model. However, instead of directly performing adversarial training, SecAlign phrases the task of defending against prompt injections in terms of preference optimization. It aims to train the model to prefer to follow instructions coming from trusted sources over untrusted sources using Direct Preference Optimization (DPO). SecAlign is a stronger defense than StruQ - in addition to blocking $\approx 100\%$ of handcrafted attacks, it also blocks $\approx 100\%$ of automated attacks crafted using previous SOTA algorithms like GCG and AdvPrompter ~\cite{zou2023universaltransferableadversarialattacks, paulus2025advprompterfastadaptiveadversarial}.

\noindent\textbf{SecAlign++} is an example of an Instruction Hierarchy (IH) model. ~\cite{wallace2024instructionhierarchytrainingllms}. IH models generalize to multiple hierarchical levels of trust, system (higher trust) and user (lower trust), LLM, and Tool Output (lowest trust) as opposed to models like StruQ and SecAlign that only offer two levels: trusted and untrusted. IH models are fine-tuned to prioritize following instructions from higher trust sources over instructions from lower trust sources. Meta's SecAlign++, in particular, is an open-weights production-grade model implementing IH with three trust levels: ``system'' (higher trust), ``user'' (lower trust) and ``input'' (untrusted).

\subsection{Why does GCG not break SecAlign?}

\begin{figure*}[htbp]
\centering
\subfloat[Mean = $0.7$, Std. Dev = $8.4$]{
    \begin{minipage}{0.29\textwidth}
        \centering
        \begin{tikzpicture}
            \begin{axis}[
                width=\linewidth,
                height=4cm,
                xlabel={$D_r$},
                ylabel={Frequency},
                ybar,
                bar width=8pt,
                enlarge x limits=0.1,
                grid=major,
                grid style={dashed,gray!30},
                title={$r=1$},
                ymin=0,
                ymax=20  
            ]
            
            % Histogram for column 1
            \addplot[
                color=blue,
                fill=blue!20,
                hist={
                    bins=30,
                    data min=-30,
                    data max=30,
                }
            ] table[y index=0, col sep=comma] {data_raw/r_wise_Dr.csv};
            
            \end{axis}
        \end{tikzpicture}
    \end{minipage}
}
\hfill
\subfloat[Mean = $-0.1$, Std. Dev = $3.5$]{
    \begin{minipage}{0.29\textwidth}
        \centering
        \begin{tikzpicture}
            \begin{axis}[
                width=\linewidth,
                height=4cm,
                xlabel={$D_r$},
                ylabel={Frequency},
                ybar,
                bar width=8pt,
                enlarge x limits=0.1,
                legend pos=north east,
                grid=major,
                grid style={dashed,gray!30},
                title={$r=3$},
                ymin=0,
                ymax=20
            ]
            
            % Histogram for column 3
            \addplot[
                color=blue,
                fill=blue!20,
                hist={
                    bins=30,
                    data min=-30,
                    data max=30,
                }
            ] table[y index=2, col sep=comma] {data_raw/r_wise_Dr.csv};
            
            \end{axis}
        \end{tikzpicture}
    \end{minipage}
}
\hfill
\subfloat[Mean = $-0.2$, Std. Dev = $2.5$]{
    \begin{minipage}{0.29\textwidth}
        \centering
        \begin{tikzpicture}
            \begin{axis}[
                width=\linewidth,
                height=4cm,
                xlabel={$D_r$},
                ylabel={Frequency},
                ybar,
                bar width=8pt,
                enlarge x limits=0.1,
                legend pos=north east,
                grid=major,
                grid style={dashed,gray!30},
                title={$r=5$},
                ymin=0,
                ymax=20,
            ]
            
            % Histogram for column 5
            \addplot[
                color=blue,
                fill=blue!20,
                hist={
                    bins=30,
                    data min=-30,
                    data max=30,
                }
            ] table[y index=4, col sep=comma] {data_raw/r_wise_Dr.csv};
            
            \end{axis}
        \end{tikzpicture}
    \end{minipage}
}
\caption{Difference between guided and unguided optimization concentrates close to $0$ when appropriately averaged}
\label{fig:three_histograms}
\end{figure*}

\begin{figure}
    \centering
    \begin{minipage}{\columnwidth}
        \centering
        \begin{tikzpicture}
            \begin{axis}[
                width=\columnwidth,
                height=4.5cm,
                xlabel={Iterations},
                ylabel={Target Logprobs},
                enlarge x limits=0.1,
                legend pos=north east,
                grid=major,
                ymin=0,
                ymax=32,
                grid style={dashed,gray!30},
            ]
            
            \addplot[
                color=blue,
            ] table[x expr=\coordindex, y=guided, col sep=comma] {data_raw/comp_loss_curves.csv};

            \addplot[
                color=red,
            ] table[x expr=\coordindex, y=unguided, col sep=comma] {data_raw/comp_loss_curves.csv};

            \legend{Guided (GCG), Unguided}
            
            \end{axis}
        \end{tikzpicture}
    \end{minipage}
    \caption{Average loss curves for guided and unguided optimization are very close to each other}
    \label{fig:guided_unguided_comp}
\end{figure}

Greedy Coordinate Gradient (GCG) \cite{zou2023universaltransferableadversarialattacks} is a state-of-the-art algorithm in white-box input optimization for an LLM. The algorithm was originally proposed in the context of jailbreaking. However, later work has modified and adapted the algorithm for generating prompt injection attacks under a variety of threat models and use cases ~\cite{fu2024impromptertrickingllmagents,liu2024automaticuniversalpromptinjection,chen2025secaligndefendingpromptinjection}.

While subsequent works have used GCG with a variety of different objective functions ~\cite{fu2024impromptertrickingllmagents,liu2024automaticuniversalpromptinjection}, here we focus on GCG as an optimization algorithm for the simplest objective function - that is, as an optimization algorithm to optimize a subset of the input prompt (keeping the rest of the input prompt fixed) in order to maximize the probability of the LLM generating the attacker-chosen target string, or equivalently, minimizing the ``target logprobs'' (\cref{eq:target_logprobs}).

Furthermore, several recent works have also proposed several algorithmic enhancements to GCG ~\cite{jia2024improvedtechniquesoptimizationbasedjailbreaking,zhang2025boostingjailbreakattackmomentum,zhao2024acceleratinggreedycoordinategradient,liao2024amplegcglearninguniversaltransferable}. These works either apply transferability or aid the search procedure via better exploiting the breadth/depth trade-off during optimization or exploiting increased randomness at each step. Fundamentally, all these works still try to minimize the target logprobs by directly using the gradients of the logprobs (at a given point) as a signal to guide the optimization. In this subsection, we aim to establish empirically that the gradients of the target logprobs when used as a proxy for the actual value of the function itself \textit{don't} provide significant value to the GCG procedure. In other words, we argue empirically that the target logprobs function provides a poor landscape for gradient-guided optimization procedures like GCG.

Concretely, given an input prompt, represented as a sequence of tokens $\mathbf{x} = (x_1, x_2, \dots, x_n)$, let $I \subseteq \{1, 2, \dots n\}$ represent the indices of the tokens which can be modified by the optimization algorithm (with the rest of the tokens kept fixed). Let $\mathbf{y}_{T} = (y_1, y_2, \dots y_m)$ represent an attacker-chosen target sequence of tokens. The attacker in this case wants the LLM to output the target tokens $\mathbf{y}_{T}$ with high probability. 

The optimization problem to be solved in this case is - 
\begin{equation}
    \mathbf{x}_{I}^{*} = \underset{\mathbf{x}_{I} \in V^{|I|}} {\mathrm{argmax}}\ \Pi_{j=1}^{m} P(y_{j} \vert x_1, \dots x_n, y_1, \dots y_{j - 1}; \Theta)
\end{equation}
That is, the GCG function tries to minimize the following loss function (using the notation $\Vert$ to denote concatenation of sequences of tokens)
\begin{equation}
    \mathrm{TargetLogprobs}(\mathbf{x}, \mathbf{y}_T) = - \sum_{j=1}^m \mathrm{log}\ P(y_j \mid \mathbf{x} \Vert \mathbf{y}_{1:j-1} ; \Theta)
    \label{eq:target_logprobs}
\end{equation}
The above optimization problem is a hard discrete optimization problem for which a direct search is not feasible.

GCG addresses this with an iterative, guided, greedy search algorithm. That is, it starts with some initial sequence of tokens $\mathbf{x}^{(0)} \in V^{|I|}$ and iteratively obtains a sequence of points $\{\mathbf{x}^{(0)}, \mathbf{x}^{(1)}, \dots \} \subseteq V^{|I|}$ until some stopping criterion is reached. At the $k+1$-th iteration, GCG performs the steps - 
\begin{enumerate}[itemsep=0pt, parsep=0pt, topsep=0pt, partopsep=0pt]
\item \textbf{(Guiding step)} - A coarse filtering step which uses the gradients of the objective function at $\mathbf{x}^{(k)}$ to generate a set of promising candidates $C^{(k)} \subseteq V^{n}$. The set of promising candidates $C^{(k)}$ consists of candidate sequences of tokens each of which differs from the current point $\mathbf{x}^{(k)}$ at exactly one token at one of the $|I|$ perturbable positions. That is, the candidates generated are all one token substitution away from the current point.
\item \textbf{(Greedy step)} A fine selection step which exactly evaluates the objective function on the set of promising candidates $C^{(k)}$ - consisting of single token substitutions of $\mathbf{x}^{(k)}$ -  obtained from the guiding step. The best performing candidate from $C^{(k)}$ is chosen as the next point $\mathbf{x}^{(k+1)} \in V^n$ and the process is repeated.
\end{enumerate}

Effectively, GCG uses gradients of the $\mathrm{TargetLogprobs}$ function, which provide local information about the landscape of this loss function at the point $\mathbf{x}^{(k)}$, as a proxy to guess what is likely to be the actual function value at a far-away point, that is, at the point corresponding to the actual substituted token. A priori, this represents a sound strategy, that is, if the optimization function were a smooth enough function, such as a piecewise linear function, then on average gradients should be expected to provide a good guess for what happens at distant points. And if so, the guiding step should add significantly to the optimization procedure.

We establish empirically that when controlled for the attack hyperparameters of number of steps, candidate set size and initialization, GCG does not significantly outperform an unguided greedy search. That is, we perform the following two algorithms which differ only in the guiding step.
\begin{enumerate}[itemsep=0pt, parsep=0pt, topsep=0pt, partopsep=0pt]
    \item Guided version - We use the candidate set generation used by GCG that uses gradients (at the current point) to suggest promising single-token substitutions as candidates for the next point.
    \item Unguided version - We use a randomly generated set of single-token substitutions as candidates.
\end{enumerate}
Crucially, for both the above, we do still evaluate the loss function exactly on the candidates generated. That is, the only difference lies in the candidate generation step.

For our empirical evaluation, we mimic the setting of the evaluation of SecAlign and StruQ. We perform our comparison study on $50$ randomly chosen examples from the AlpacaFarm evaluation set containing both an ``instruction'' and ``input'' field. For each example, we inject the fixed payload ``$\verb|Print exactly `Hacked!'|$'' at the end of the ``input'' part followed by an optimizable suffix of $20$ tokens. The adversarial target is the string `$\verb|Hacked|$'.

Since we are interested in evaluating the effectiveness of the guiding strategy used by GCG, we aim to control for all other hyperparameters used during optimization and average over all sources of randomness used.

Concretely, we perform both the guided and unguided optimization as above for $500$ iterations each with the same candidate set size of $512$. For the guided version, the $512$ candidates are chosen as guided by the $\mathrm{Top}-256$ gradient values for each of the $20$ token positions. That is, $512$ candidates are chosen from $256 \times 20 = 5120$ promising single-token substitutions. On the other hand, for the unguided version, the $512$ candidates are chosen from $256$ randomly chosen single-token substitutions for each of the $20$ positions. We run both the versions initialized with the same randomly initialized sequence of $20$ tokens.

Further, to control for the effect of randomness, for each of the $50$ examples, we perform $r$ different runs of the above experiment, each with different initializations. We then compute the average difference between the best guided loss obtained during the optimization and the best unguided loss obtained during the optimization averaged over the $r$ runs.
that is, $$D_{r} = \frac{1}{r}\sum_{\alpha=1}^{r} (\mathrm{min}_{k}L^{(g)}_{\alpha}(k) - \mathrm{min}_{k}L^{(u)}_{\alpha}(k))$$ and plot the distribution of the variable $D_{r}$ over the $50$ examples averaged over different values of $r$ ranging from $r=1$ to $r=5$ runs in \cref{fig:three_histograms}. The distributions of $D_r$ for different values of $r$ establish the following -

\textbf{Across examples, guided optimization doesn't significantly outperform unguided.}
For all values of $r$, $D_r$ is distributed approximately symmetrically around $0$. That is, the guided version outperforms the unguided by about as much as the unguided version outperforms the guided one. This can be seen by the average of $D_r$ (averaged across examples, for a fixed $r$), which is close to $0$ for all values of $r$.
The same behaviour is observed over the course of the optimization procedure as is seen in the average loss curves for the guided and unguided optimizations in \cref{fig:three_histograms}.
% This is fundamentally since the objective function used by GCG - \textit{i.e.}, the objective function that directly maximizes the probability of target tokens - provides a noisy landscape for optimization. We show an example of a slice of the landscape in \cref{a}. As can be seen, the value of the objective function along the path joining these points fluctuates significantly along the path. These fluctuations in the landscape significantly hamper the usability of gradients as a useful signal for the optimization procedure. That is, for this loss function, gradients are a poor predictor for what happens far away from the current point.

\textbf{The difference between the performance of guided and unguided optimizations depends strongly on the initialization and other randomness.}
We observe that with increase in $r$, the variance and spread in the distribution of $D_r$ (across examples) decreases rapidly (\cref{fig:three_histograms}). That is, any outsized difference observed between the performance of the guided and unguided versions of the optimization on a particular single run (for a single example) disappears when more runs are averaged for the same example. In particular, this means that the role played by the initializer and randomness are crucial to ensure that GCG outperform an unguided search.

From the above, we establish empirically that the loss landscape provided by the loss function $\mathrm{TargetLogprobs}$ represents a poor landscape for optimization and makes it so that GCG can't utilize gradients of $\mathrm{TargetLogprobs}$ with good effect.
\section{The \att\ Attack}
\label{sec:attack}

To address the above problems, we propose a new method that explicitly exploits the internal structure of an LLM. Concretely, we use the architectural properties of LLMs in order to craft a novel loss function that can be used to generate better prompt injection attacks. We use an optimized input generated by optimizing this loss function as a ``warm start'' initializer to ``fine-tune'' the adversarial input using standard algorithms like GCG.

Our key insight is that we can exploit an instruction-tuned model's instruction-following ability to craft prompt injection attacks. Specifically, we note that if the attacker's instruction payload was the only text in the input stream, the model would easily follow that instruction. This insight forms the basis for our loss function. To formalize this mathematically, we rely on the attention mechanism of an LLM. The attention mechanism forms the key component of our new loss function.

\subsection{Background on Attention}

The attention mechanism in an LLM is a core component which assigns a weight to each previous token while trying to generate the next token.

Formally, when a sequence of tokens $\mathbf{x} = (x_1, x_2, \dots x_n)$ is inputted into an LLM, it first maps each token $x_i$ to an embedding vector $e_i^{(0)} \in \mathbb{R}^{d}$ to obtain a sequence of initial representations $(e_1^{(0)}, e_2^{(0)}, \dots e_n^{(0)}) \in \mathbb{R}^{n \times d}$. This sequence of representations is transformed by a series of decoder layers (say, total of $L$ decoder layers) into a final transformed representation $(f_1, f_2, \dots f_n) \in \mathbb{R}^{n \times d}$. Finally, the ``language modelling head", represented by a matrix $W \in \mathbb{R}^{d \times |V|}$ is used to generate the probability distribution for the next token by the following calculation - 
$$P(\cdot \vert x_1, x_2, \dots x_n; \Theta) = \mathrm{softmax}(f_n \cdot W) \in \mathbb{R}^{|V|}$$ where each coordinate lies between $0$ and $1$ and represents the probability of generation of that particular token given the context $(x_1, x_2, \dots x_n)$.

The $l$-th decoder layer $D^{(l)}$ is itself a complex computation that maps one sequence of representations in $\mathbb{R}^{n \times d}$ to another sequence of representations in $\mathbb{R}^{n \times d}$. It consists of an attention layer (typically prenormalized and postnormalized) and a fully connected layer that modify a residual stream that forms the internal representations that feed into the next decoder layer $D^{(l+1)}$. For the sake of brevity, here we only describe some relevant specifics of the attention mechanism. A rigorous and complete treatment of transformers can be found in \cite{phuong2022formalalgorithmstransformers}.

The key architectural innovation at the core of a decoder layer is the Masked Multi-head Attention (MMHA) mechanism. The MMHA mechanism at the $l$-th decoder layer $D^{(l)}$ takes in a sequence of representations obtained from the previous layer, say $E^{(l-1)} = (e_1^{(l-1)}, e_2^{(l-1)}, \dots, e_n^{(l-1)}) \in \mathbb{R}^{n \times d}$ and computes a sequence of square matrices $(A_1^{(l)}, A_2^{(l)}, \dots A_{H}^{(l)})$ called ``attention matrices'' where $A_i^{(l)} \in \mathbb{R}^{n \times n}$. The number $H$ represents the number of ``attention heads'' of the LLM at layer $l$.
These matrices are computed as
\begin{equation}
    A_i^{(l)} = \mathrm{softmax}(E^{(l-1)}W_i^{(l)}E^{(l-1)T} + \mathrm{CausalMask}_{n})
    \label{eq:att_matrix}
\end{equation}

Here, $W_i^{(l)} \in \mathbb{R}^{d \times d}$ is a pre-trained parameter of the LLM corresponding to the $i$-th attention head at the $l$-th layer. The matrix $\mathrm{CausalMask}_n$ is an $n \times n$ matrix with $-\infty$ above the diagonal and $0$ everywhere else, and the $\mathrm{softmax}$ function is applied row-wise to the matrix.
The attention matrices $A_i^{(l)}$ represent intermediate values computed within the decoder layer $D^{(l)}$ that are ultimately used to further calculate the representations that can be used as input for the next layer $D^{(l+1)}$.

Due to the specific functional form above, an attention matrix $A$ has significant structure to it. All attention matrices have the following properties. (We use the notation $A[j]$ to refer to the $j$-th row of $A$ and the notation $A[j][k]$ to refer to the element in the $j$-th row and $k$-th column of $A$)

% \al{I think it most importantly should state an intuition what $A[j][k]$ means (attention from one token to the other)}
% \np{It is written in some more detail in the subsequent enumerate}

\begin{enumerate}[itemsep=0pt, parsep=0pt, topsep=0pt, partopsep=0pt]
    \item The attention matrices are lower-triangular, \textit{i.e.}, elements above the diagonal are zero. That is, $A[j][k] = 0$ if $j < k$.
    \item Each element of the attention matrix lies between $0$ and $1$ (inclusive). That is, $0 \leq A[j][k] \leq 1$
    \item Each row of the attention matrix sums up to $1$. That is, for each $j$,
    $\sum_{k=1}^{n}A[j][k] = \sum_{k=1}^{j}A[j][k] = 1$
\end{enumerate}

The above properties imply the following interpretation(s) for the rows of an attention matrix $A$. 
\begin{enumerate}[itemsep=0pt, parsep=0pt, topsep=0pt, partopsep=0pt]
    \item The first $j$ entries in the $j$-th row $A[j]$ of an attention matrix represent the attention given (by that head) to the first $j$ tokens when generating the $j+1$-th token.
    \item The first $j$ entries in the $j$-th row are the weights of a convex combination over $j$ dimensions
    \item The first $j$ entries in $A[j]$ are the probability masses of a discrete probability distribution over \{1, 2, \dots j\}
\end{enumerate}

\begin{figure}[t]
\centering
\[
\begin{array}{r@{\hskip 1pt}c}
  % Row labels on the left
  \begin{array}{r}
    \textit{``What"} \\
    \textit{``is"} \\
    \textit{``the"} \\
    \textit{$\vdots$} \\
    \textit{``+"} \\
    \textit{``2?"} \\
    \\
  \end{array}
  &
  % Matrix with big parentheses
  \left(
  \begin{array}{ccccc}
    1 & 0 & \cdots & 0 & 0 \\
    0.3 & 0.7 & \cdots & 0 & 0 \\
    0.2 & 0.3 & \cdots & 0 & 0 \\
    \vdots & \vdots & & \vdots & \vdots \\
    0.06 & 0.14 & \cdots & 0.03 & 0  \\    
    0.05 & 0.1 & \cdots & 0.09 & 0.01  \\
    \hline
    \textit{``What"} & \textit{``is"} &
    \textit{$\cdots$} &
    \textit{``+"} &
    \textit{``2?"} \\
  \end{array}
  \right)
  \end{array}
\]

\caption{An attention matrix for the input ``What is the value of 2 + 2?". The last row contains the weights assigned to each of the previous tokens for generating the next token.}
\label{fig:ex_att_mat}
\end{figure}

\subsection{Crafting a loss function}

We use the above interpretations to guide the design of our loss function. When the LLM is processing the $n$ tokens in the input $\mathbf{x} = (x_1, x_2, \dots x_n)$ in order to generate the $n+1$-th token, the last (i.e. $n$-th) rows of all the attention matrices $A_{i}^{(l)}(\mathbf{x})$ represent the attention given to each of the $n$ tokens in the input by that attention head. Since we want the model to attend only to the payload, we identify the indices of tokens corresonding to the attacker's payload, say $J \subseteq \{1, 2, \dots n \}$ and try to maximize the sum of attentions given to this part of the input. As before, let $\mathbf{x}_{I}$ represent the modifiable tokens which are to be optimized by the attacker. Let $y$ be the target token that the attacker wants the LLM to generate as the $n+1$-th token.

Formally, for a single attention matrix $A_{i}^{(l)}(\mathbf{x})$ generated on input $\mathbf{x} = (x_1, x_2, \dots, x_n)$ of length $n$, we define 
% \al{I think it's unclear what $n$ means - do we want attention from the last attacker's token or from the 'current next token model is about the generate'?}
% \np{Clarified with notation}
\begin{equation}
    \mathrm{AttLoss}_i^{(l)}(\mathbf{x}, y) = 1 - \sum_{j \in J}A_{i}^{(l)}(\mathbf{x})[n][j]
    \label{eq:single_att_loss}
\end{equation}
We note that $\mathrm{AttLoss}_i^{(l)}(\mathbf{x}, y)$ doesn't depend on the target token $y$ and only depends on the input.

To account for all the attention matrices, we define 
\begin{equation}
    \mathrm{AttLoss}(\mathbf{x}, y) = \sum_{l=1}^{L}\sum_{i=1}^{H}w_{i}^{(l)}\mathrm{AttLoss}_{i}^{(l)}(\mathbf{x}, y)
    \label{eq:combined_att_loss}
\end{equation}
where $\{w_i^{(l)}\}$ are some set of weights, assigned to all the attention heads.

For a longer sequence of target tokens $\mathbf{y} = (y_1, y_2, \dots, y_m)$, we just sum up the losses over the target tokens and define
\begin{equation}
    \mathrm{AttLoss}(\mathbf{x}, \mathbf{y}) = \sum_{j=1}^{m} \mathrm{AttLoss}(\mathbf{x} \Vert \mathbf{y}_{1:j-1}, y_j)
    \label{eq:long_combined_att_loss}
\end{equation}

This loss function becomes $0$ when the entire attention of each attention matrix is concentrated on the attacker payload tokens. That is, this loss function captures how much of the attention of the LLM is being paid to the surrounding context. A larger value of this loss function implies that more of the attention across all layers is distributed over the entire context window, whereas a smaller value implies that a lesser amount of attention is distributed over the entire context.

\subsection{Weights for each attention head}

\begin{figure}[t]
\centering
\begin{subfigure}[b]{0.40\textwidth}
    \includegraphics[width=\textwidth]{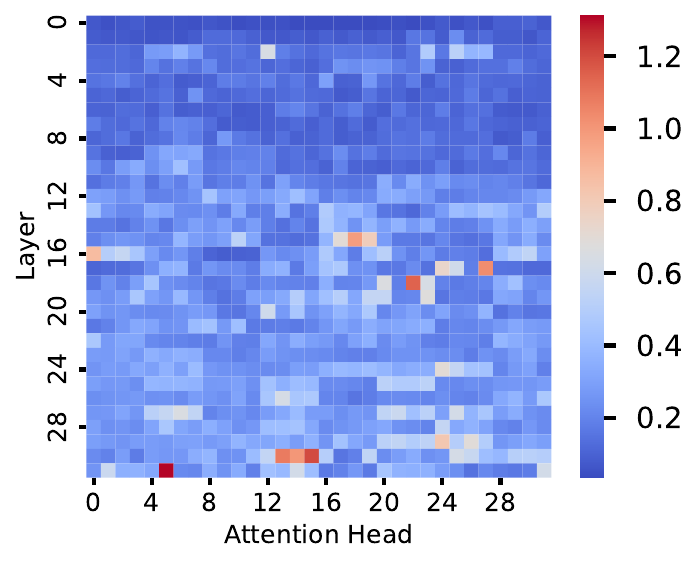}
    \caption{SecAlign Meta-Llama-3-8B-Instruct}
    \label{fig:meta_sens}
\end{subfigure}
\hfill
\begin{subfigure}[b]{0.40\textwidth}
    \includegraphics[width=\textwidth]{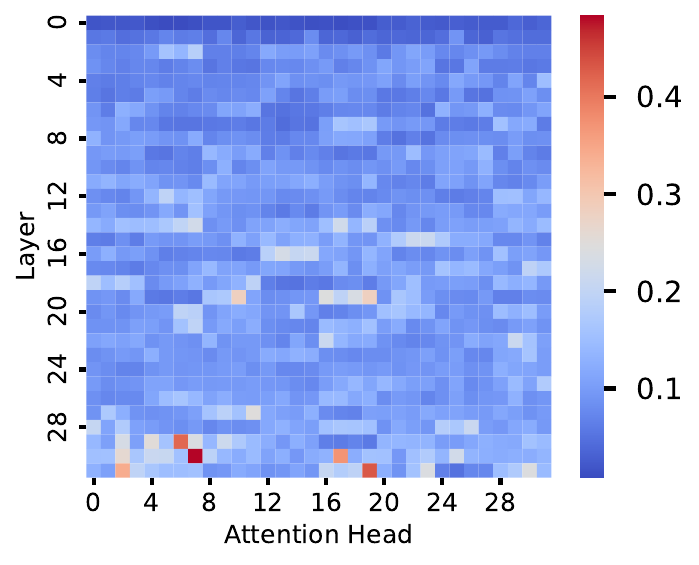}
    \caption{SecAlign Mistral-7B-v0.1-Instruct}
    \label{fig:mistral_sens}
\end{subfigure}
\caption{Very few attention heads have high sensitivities for the target string ``Hacked''}
\label{fig:heatmaps}
\end{figure}

In principle, we could simply add all contributions equally from each attention head $\mathrm{AttLoss}_{i}^{(l)}$ to craft the loss function $\mathrm{AttLoss}$. However, this is likely to be suboptimal since 
\begin{enumerate}[itemsep=0pt, parsep=0pt, topsep=0pt, partopsep=0pt]
    \item Prior work from the field of mechanistic interpretability has shown that different attention heads carry out different functions in an LLM. ~\cite{olsson2022incontextlearninginductionheads,yin2025attentionheadsmatterincontext}
    \item Attention heads across layers are not independent from each other, since the value of the attention matrix in the $l+1$-th layer depends on the value of the attention matrix in the $l$-th layer. Small changes in the values of the attention matrix in earlier layers can lead to large changes in the values of attention matrices in later layers.
\end{enumerate}

% \al{Ok, I think this is where it doesn't match my intuition. Prior works in mechanistic interpretability shows that only a few attention heads are important, but it achieves result by actually acting on all of them. It shows that existing adversarial prefixes mostly act on those heads, but it doesn't say that acting on others will invalidate the result.
% We can instead say that we leverage previous fact that sometimes some attention heads are more important and therefore incorporate the weights.}
% \np{Yes, that is what we are going for. Changed the phrasing a bit to reflect that}

In general, figuring out which attention head is important for which task requires extensive manual analysis and/or lengthy ablation experiments which turn off attention heads one-by-one to see the effect they have on the LLM's output ~\cite{yin2025attentionheadsmatterincontext}. A typical LLM has more than a thousand attention heads (across all layers) which makes such an analysis impractical.

Our key insight is that since our eventual goal is finding adversarial perturbations in the token space that lead to an adversary chosen output, we don't need to explicitly find out which attention heads are functionally important for certain abstract tasks. Instead, we only need to find those attention heads for which small changes in their corresponding attention matrices lead to large changes in the probability of the desired target token. This observation allows us to sidestep an expensive ablation experiment which would turn off attention heads one-by-one and analyze its impact on the output. Instead, since sensitivity to small perturbations (of the probability of the target token) is a local property of a function, sensitivity can be meaningfully quantified using gradients. We use these gradients as the key piece of information that allows us to quantify the importance of attention heads.

Formally, let $y$ be a target token that the adversary wants the LLM to generate. Let $\mathbf{z} = (z_1, z_2, \dots z_n)$ represent an input sequence of tokens which generate the attention matrices $\{A_i^{(l)}(\mathbf{z})\}$ during the forward pass of the LLM.
Let $$\mathrm{LogProb}(y \vert \mathbf{z}) = \mathrm{log}P(y \vert \mathbf{z}; \Theta)$$ represent the log of the probability of the target token $y$. Then, $$\frac{\partial\ \mathrm{LogProb}(y\vert \mathbf{z})}{\partial\ A_{i}^{(l)}(\mathbf{z})} \in \mathbb{R}^{n \times n}$$ represents the gradients of this loss function with respect to the attention matrices at the input point $\mathbf{z}$. Since the $n+1$-th token prediction depends on the attention values in the $n$-th row,  we look at the sum of the magnitude of the entries in the $n$-th row of the gradient matrix to get a measure for how sensitive the loss function is to changes in that row of the matrix.

That is, we define sensitivity as
\begin{equation}
    \mathrm{Sen}_{i}^{(l)}(\mathbf{z}, y) = \sum_{k=1}^{|\mathbf{z}|}\left| \frac{\partial\ \mathrm{LogProb}(y \vert \mathbf{z})}{\partial\ A_{i}^{(l)}(\mathbf{z})}[n][k]\right|
    \label{eq:sens_sing_att}
\end{equation}

For a longer target sequence $\mathbf{y} = (y_1, y_2, \dots y_m)$, we define
% need to account for all the attention matrices that occur during the autoregressive generation of the next $m$ tokens on the input sequence $\mathbf{z}$. That is, after generating $j-1$ output tokens (\textit{i.e.} $y_1, y_2, \dots y_{j-1}$), the attention matrix 

\begin{equation}
    \mathrm{Sen}_i^{(l)}(\mathbf{z}, \mathbf{y}) = \sum_{j=1}^{|\mathbf{y}|} \mathrm{Sen}_i^{(l)}(\mathbf{z}\Vert \mathbf{y}_{1:j-1}, y_j)
    \label{eq:sens_longer}
\end{equation}

Finally, to get an average sensitivity score for an attention head (with respect to a target sequence of tokens $\mathbf{y}$), we take a dataset $\mathcal{D}$ of inputs $\mathbf{z}$ and normalize the sensitivities by the total length $|\mathbf{z}| + |\mathbf{y}|$ and average out over the dataset $\mathcal{D}$
\begin{equation}
    \overline{\mathrm{Sen}}_i^{(l)}(\mathbf{y}) = \frac{1}{|\mathcal{D}|}\sum_{\mathbf{z} \in \mathcal{D}} \frac{1}{|\mathbf{z}| + |\mathbf{y}|}\mathrm{Sen}_i^{(l)}(\mathbf{z}, \mathbf{y})
    \label{eq:avg_sens}
\end{equation}

% \al{why do we measure sensitivity (gradients magnitude) w.r.t to the target output loss and not w.r.t to the attention values loss? what would happen if we don't use sensitivity later in the final attention loss?}
% \np{The loss values are lower with weights}

% \al{sensitivity loss equation uses l1 norm of gradients, which is also used for out-of-distribution detection approach. Shall we need an explanation/discussion of that?}
% \np{No, IMO}

This provides a proxy measure for how sensitive, on average, the probability of a target token $y$ is to changes in the value of the attention matrix at the $i$-th attention head at the $l$-th layer. We show the average sensitivities for two different SecAlign defended models for the target string ``$\verb|Hacked|$'' constructed from a training dataset consisting of examples from the Dolly fine-tuning dataset ~\cite{DatabricksBlog2023DollyV2} in \cref{fig:heatmaps}.

In practice, we find that a large subset of the sensitivities (across heads and layers) are very small, as can be seen in \cref{fig:heatmaps}. Heuristically, small changes in these attention heads don't lead to large changes in the probability of the target sequence of tokens and are likely to act as noise during optimization. We can therefore choose a threshold $\tau$ and manually set the small contributions of heads less than $\tau$ to $0$. That is, we define the clipped sensitivities as 
\begin{equation}
    \mathrm{ClipSen}_i^{(l)}(\mathbf{y}) =
    \begin{cases}
        0 & \mathrm{if}\  \overline{\mathrm{Sen}}_i^{(l)}(\mathbf{y}) \leq \tau \\
        \overline{\mathrm{Sen}}_i^{(l)}(\mathbf{y}) & \mathrm{otherwise}  \\
    \end{cases}
    \label{eq:clip_sen}
\end{equation}

\begin{algorithm}[t]
    \SetAlgoLined
    \KwIn{Initial sequence of tokens $\mathbf{x}^{(0)} = (x_1, x_2, \dots, x_n) \in V^{n}$, with modifiable subset of indices $I \subseteq \{1, 2, \dots, n\}$}
    \KwIn{Target tokens $\mathbf{y}_T = (y_1, y_2, \dots, y_m) \in V^{m}$}
    \KwIn{A function $\mathrm{Loss}: V^{n} \times V^{m} \to \mathbb{R}$ to be optimized}
    \textbf{Parameters: } $p \in \mathbb{N}$ — number of top candidates per token position at each iteration; 
    $N \in \mathbb{N}$ — number of iterations; 
    $B \in \mathbb{N}$ — number of function evaluations per iteration. \\
    \KwOut{Optimized tokens $\mathbf{x}^{*}$}

    \Begin{
        Logprobs $\leftarrow [\ ]$ \;
        Tokens $\leftarrow [\ ]$ \;

        \For{$k \leftarrow 0$ \KwTo $N-1$}{
            \ForEach{$i \in I$}{
                $X_i^{(k)} \leftarrow \mathrm{Top\text{-}p}\left( -\nabla_{e_{x_i}} \mathrm{Loss}(\mathbf{x}^{(k)}, \mathbf{y}_T) \right)$
            }

            \For{$b \leftarrow 1$ \KwTo $B$}{
                $i \sim \mathrm{Uniform}(I)$ \\
                $\tilde{\mathbf{x}}^{(k)}_{b} \leftarrow \mathbf{x}^{(k)}$ \\
                $\tilde{\mathbf{x}}^{(k)}_b[i] \sim \mathrm{Uniform}(X_i^{(k)})$
            }

            $b^* \leftarrow \arg\min_b\ \mathrm{Loss}(\tilde{\mathbf{x}}_{b}^{(k)})$ \\
            $\mathbf{x}^{(k+1)} \leftarrow \tilde{\mathbf{x}}^{(k)}_{b^{*}}$\\
            Logprobs$[k] \leftarrow \mathrm{TargetLogprobs}(\mathbf{x}^{(k+1)}, \mathbf{y}_T)$\\
            Tokens$[k] \leftarrow \mathbf{x}^{(k+1)}$
        }

        \Return Tokens[argmin Logprobs]
    }
    \caption{\texttt{\textbf{GenGCG}}$_{p, N, B}(\mathbf{x}^{(0)}, \mathbf{y}_T, \mathrm{Loss})$}
    \label{alg:custom_gcg}
\end{algorithm}
\begin{algorithm}[!h]
    \SetAlgoLined

    \KwIn{Initial sequence of tokens $\mathbf{x}^{(0)} = (x_1, x_2, \dots, x_n) \in V^{n}$, with modifiable subset of indices $I \subseteq \{1, 2, \dots, n\}$}
    \KwIn{Target tokens $\mathbf{y}_T = (y_1, y_2, \dots, y_m) \in V^{m}$}
    \textbf{Parameters: } $p \in \mathbb{N}$ — number of top candidates to consider; 
    $N_1 \in \mathbb{N}$ — number of steps for first phase; 
    $N_2 \in \mathbb{N}$ — number of steps for second phase; 
    $B \in \mathbb{N}$ — number of evaluations per step. \\
    \KwOut{Optimized tokens $\mathbf{x}^{*}$}

    \Begin{
        $\mathbf{x}_1^{*} \leftarrow \texttt{\textbf{GenGCG}}_{p, N_1, B}(\mathbf{x}^{(0)}, \mathbf{y}_T, \mathrm{AttLoss})$ \\
        $\mathbf{x}_2^{*} \leftarrow \texttt{\textbf{GenGCG}}_{p, N_2, B}(\mathbf{x}_1^{*}, \mathbf{y}_T, \mathrm{TargetLogprobs})$ \\
        \Return $\mathbf{x}_2^{*}$
        }

    \caption{$\texttt{\textbf{\att}}_{p, N_1, N_2, B}(\mathbf{x}^{(0)}, \mathbf{y}_T)$}
    \label{alg:final_attack_alg}
\end{algorithm}

We use these clipped sensitivities as the final weights for our attention loss function to get
\begin{equation}
    \mathrm{AttLoss}(\mathbf{x}, \mathbf{y}) = \sum_{l=1}^{L}\sum_{i=1}^{H} \mathrm{ClipSen}_i^{(l)}(\mathbf{y}) \cdot \mathrm{AttLoss}_i^{(l)}(\mathbf{x}, \mathbf{y})
    \label{eq:att_loss_sing_token}
\end{equation}

\subsection{~\att\ : Strong Knowledge Adversary}

We first begin by describing our final attack for the simpler case of a Strong Knowledge Adversary (\cref{sec:threat_model}) who has full knowledge of the exact conversation context into which the attack is placed. In this particular case, the adversary has to optimize for a target output over a single input prompt. That is, the goal of the adversary is to find an input $x_{Adv}$ such that $LLM_{\Theta}(\mathrm{Combine}(x_{Trusted}, x_{Adv})) \approx y_{Target}$ (\cref{eq:adv_goal_pi}) for the special case when the adversary knows the string $x_{Trusted}$. This represents a simple prompt injection objective from an optimization point of view.

~\att\ is a two-phase attack where instead of directly optimizing for the $\mathrm{TargetLogprobs}$ objective, we instead begin by minimizing the attention loss function $\mathrm{AttLoss}$ (\cref{eq:att_loss_sing_token}) to get a ``warm start'', that is, a good initialization that can be ``fine-tuned'' by then minimizing for $\mathrm{TargetLogprobs}$. To compute the attention loss function, we use a proxy dataset such as the Dolly fine-tuning dataset to calculate the sensitivities and subsequently the exact loss values ~\cite{DatabricksBlog2023DollyV2}.

To solve the actual optimization problems in both phases, we use essentially the same guided greedy approach as used by the GCG algorithm, with the only change being the loss function under consideration. That is, we rely on the gradients at the current point to provide us with a candidate set of promising single-token substitutions and we evaluate the loss exactly on these substitutions to get the next point in the process. We call this algorithm Generalized GCG (\cref{alg:custom_gcg}). Our final two-phase algorithm is as described in \cref{alg:final_attack_alg}.

% \al{I actually think, how does it look through the prism of a usual chain rule if we expand the LogProb(y|z) loss function? This new loss looks like two times there is an expectation of LogProb(y|z) taken over input tokens, and some terms (or coefficients) in the final chain rule are replaced with some averages. Then, taking gradients from this function at the guiding GCG step should result in the similar problem we are trying to address?}

\begin{figure*}[]
\centering
\begin{subfigure}[b]{0.33\textwidth}
    \includegraphics[width=\textwidth]{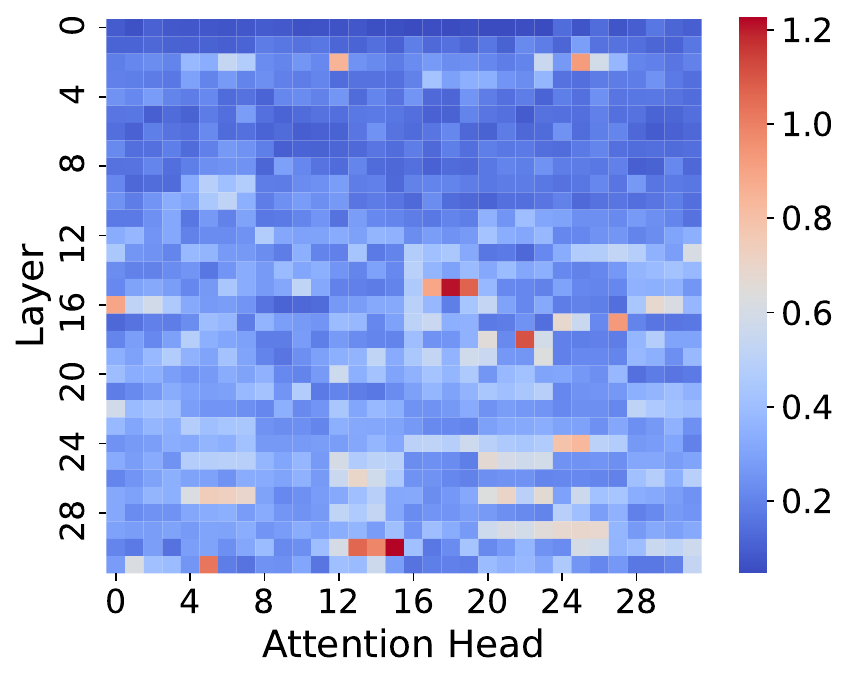}
    \caption{Step 0 \textit{(Avg. Target Logprobs: 26.3)}}
    \label{fig:sens_0}
\end{subfigure}
\hfill
\begin{subfigure}[b]{0.33\textwidth}
    \includegraphics[width=\textwidth]{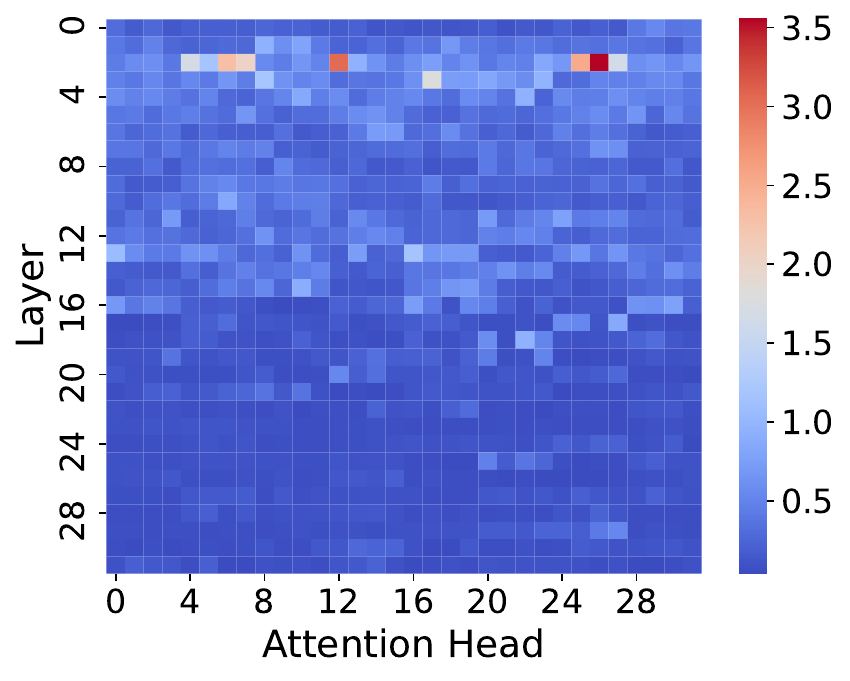}
    \caption{Step 200 \textit{(Avg. Target Logprobs: 3.52)}}
    \label{fig:sens_1}
\end{subfigure}
\hfill
\begin{subfigure}[b]{0.33\textwidth}
    \includegraphics[width=\textwidth]{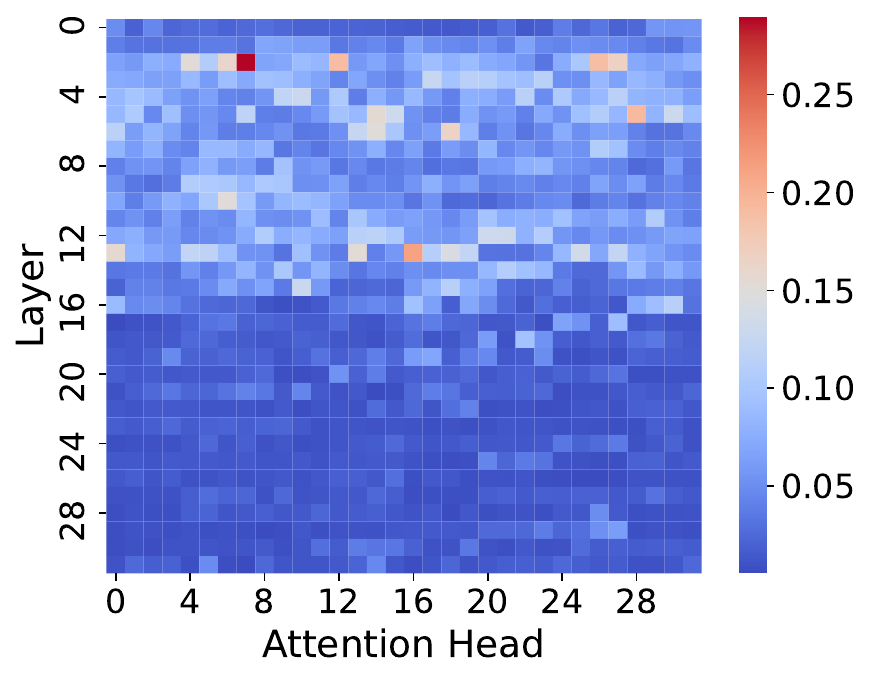}
    \caption{Step 400 \textit{(Avg. Target Logprobs: 0.23)}}
    \label{fig:sens_2}
\end{subfigure}
\caption{Local Sensitivity patterns at points with high average target logprobs (Step 0) - differ from those at points with low average target logprobs (Steps 200 and 400). (All sensitivities obtained by averaging over a dataset of size 10.)}
\hfill
\label{fig:local_sens}
\end{figure*}

\subsection{~\att ++ : Weak Knowledge Adversary}

In contrast to a Strong Knowledge Adversary, a Weak Knowledge Adversary doesn't have any information about the surrounding context tokens $\mathbf{x}_{\mathrm{Trusted}}$. Therefore, the adversary needs to be able to generate \textit{universal} prompt injections that can work across a large range of $\mathbf{x}_{\mathrm{Trusted}}$ values. This represents a significantly harder optimization objective than in the case of a strong knowledge adversary.

To address this problem, we propose a variant of \att, which we call \att++ that dynamically updates the weights assigned to the attention heads as optimization progresses. At a high-level, this captures the idea that as the adversarial perturbations go from regions with high target logprobs (in-distribution) to regions with low target logprobs (out-of-distribution), the sensitivities, which are themselves a function of the logprobs (\cref{eq:avg_sens}), also consequently change and hence dynamically updating the attention loss function gives us a better view of which attention heads are how important at any step in the optimization process.

Prior work has addressed the task of generating universal prompt injections by attempting to optimize the adversarial perturbation sequence of tokens $\mathrm{r}$ over a training set of surrounding contexts by directly optimizing for the \textit{average} target logprobs (of the target string) over the training data ~\cite{pasquini2024neuralexeclearningand, liu2024automaticuniversalpromptinjection}. That is, they use a training dataset where each training datapoint consists of a sequence of context tokens $\mathbf{x}_{\mathrm{Trusted}}$, a sequence of prompt injection payload tokens $\mathbf{x}_{\mathrm{Payload}}$ and the appropriate target sequence of tokens $\mathbf{y}_T$ corresponding to the payload $\mathbf{x}_{\mathrm{Payload}}$. That is, $\mathcal{D} = \{(\mathbf{x}_{\mathrm{Trusted}}, \mathbf{x}_{\mathrm{Payload}}, \mathbf{y}_T)\}$. For simplicity of exposition and notation, but without loss of generality, we assume that the modifiable tokens to be optimized, say, $\mathbf{r}$, are always placed in a suffix-only configuration after the payload string. The loss function used by prior works for universality in this case is -
\begin{equation}
    \overline{\mathrm{Logprobs}}_{\mathcal{D}}(\mathbf{r}) = \underset{\mathcal{D}}{\mathbb{E}}\ \mathrm{TargetLogprobs}(\mathbf{x}_{\mathrm{Trusted}}\Vert\mathbf{x}_{\mathrm{Payload}}\Vert\mathbf{r}, \mathbf{y}_T)
    \label{eq:exp_tar_logprobs}
\end{equation}

While we adopt a similar approach for constructing our universal prompt injections, the nature of a prompt injection helps us simplify the optimization task significantly. Optimizing over several different choices of contexts, payloads and target strings represents a hard training objective. Our key insight is that, for the case of a prompt injection attack specifically, a realistic attacker needn't worry about optimizing over several payloads. Instead, since the attacker controls the payload, we can simplify the optimization task by reasonably assuming that the sequence of tokens $\mathbf{x}_{\mathrm{Payload}}$ is constant across all the training examples. Similarly, we can also assume that the target sequence $\mathbf{y}_T = \mathbf{y}$ is constant across all training examples. As a consequence, our training dataset varies only in the contents of the surrounding contexts. We use the notation $\mathcal{D} = \{\mathbf{x}_{\mathrm{Trusted}} \Vert \mathbf{x}_{\mathrm{Payload}}\}$ to refer to the training set of contexts over which the training has to occur. For this simplified case, we use the following notation - 
\begin{equation}
    \overline{\mathrm{Logprobs}}_{\mathcal{D},\mathbf{y}}(\mathbf{r}) = \underset{\mathbf{x} \in \mathcal{D}}{\mathbb{E}}\mathrm{TargetLogprobs}(\mathbf{x}\Vert\mathbf{r}, \mathbf{y})
    \label{eq:simplified_avg_target_logprobs}
\end{equation}

Analogous to the single-prompt optimization case, instead of directly minimizing the average target logprobs loss function, we try to create a ``warm start'' which can be later improved by optimizing the average target logprobs. 

We use a similar attention loss construction in order to craft our warm start with a few crucial differences. Instead of pre-computing a globally constant set of sensitivities from a fixed, proxy dataset to use as weights, we directly try to leverage the training dataset $\mathcal{D}$ to compute \textit{local} sensitivities. 

For each training datapoint $\mathbf{x} = \mathbf{x}_{\mathrm{Trusted}} \Vert \mathbf{x}_{\mathrm{Payload}} \in \mathcal{D}$ and a fixed perturbation $\mathbf{r}$, we can create the perturbed point $\mathbf{x}_{\mathrm{Trusted}}\Vert\mathbf{x}_{\mathrm{Payload}}\Vert\mathbf{r}$ and the dataset $\mathcal{D}\oplus\mathbf{r} = \{\mathbf{x}\Vert\mathbf{r}\ \mid\ \mathbf{x}\in \mathcal{D} \}$.

At a given perturbation $\mathbf{r}$, we can compute the local sensitivities (over the dataset $\mathcal{D} \oplus \mathbf{r}$) for a target sequence $\mathbf{y}$ as 
\begin{equation}
    \mathrm{LSen}_{i, \mathbf{r}}^{(l)}(\mathcal{D}, \mathbf{y}) = \frac{1}{|\mathcal{D} \oplus \mathbf{r}|} \sum_{\mathbf{z} \in \mathcal{D} \oplus \mathbf{r}} \frac{1}{|\mathbf{z}| + |\mathbf{y}|} \mathrm{Sen}_i^{(l)}(\mathbf{z}, \mathbf{y})
\end{equation}
This is the same construction as the global sensitivities (\cref{eq:avg_sens}) except that these are being computed over a dataset which is itself adversarially perturbed.
Like for \att, we clip these local sensitivities to a threshold $\tau$ to get the clipped local sensitivities, which we denote by $\mathrm{CLSen}_{i, \mathbf{r}}^{(l)}(\mathcal{D}, \mathbf{y})$ .

These local sensitivities provide a more relevant quantification of the relative importance of attention heads since they take into account the effect of the perturbation itself on the probabilities of the target tokens. In contrast, sensitivities obtained by averaging over a proxy, third-party dataset can't capture the effect of the perturbation itself since they only capture the sensitivities over a static dataset -- which only reflect ``in-distribution'' points. We find empirically that sensitivities do change significantly when going from in-distribution inputs to out-of-distribution inputs (\cref{fig:local_sens}) thus motivating our next definition.

We define the Average Local Attention Loss (with local sensitivities computed at a perturbation $\mathbf{r}$, averaged over $\mathcal{D}$) at an input point $\mathbf{x}$ and a target sequence of tokens $\mathbf{y}$ as 
\begin{equation}
    \overline{\mathrm{AttLoss}}_{\mathbf{r}, \mathcal{D}}(\mathbf{x}, \mathbf{y}) = \sum_{l=1}^{L} \sum_{i=1}^{H} \mathrm{CLSen}_{i, \mathbf{r}}^{(l)}(\mathcal{D}, \mathbf{y}) \cdot \mathrm{AttLoss}_i^{(l)}(\mathbf{x}, \mathbf{y})
    \label{eq:av_loc_att_loss}
\end{equation}

For our final algorithm \att ++, we use the same two phase approach as for \att, where in the first phase, we minimize the \textit{average} attention loss, followed by minimizing the \textit{average} target logprobs (losses averaged over the training dataset) with a key difference. While minimizing the attention loss, we periodically refresh the weights of the loss function according to where we are in the optimization process. The full algorithm is as shown in \cref{alg:astra_plus_plus} in \cref{app:astra_plus_plus}.
\section{Evaluation}
\label{sec:eval}

With our evaluation, we aim to answer the following main research questions - 

\begin{enumerate}[itemsep=0pt, parsep=0pt, topsep=0pt, partopsep=0pt]
    \item How effective is \att\ in breaking fine-tuning based defenses under the Strong Knowledge Assumption?
    \item How effective is \att ++ in breaking these defenses on unseen contexts (the Weak Knowledge Assumption)?
    \item How does our attack depend on the choice of weights?
\end{enumerate}

\subsection{Attack Success Rate - \att}

\noindent \textbf{Models.} For our evaluation, we choose the SecAlign defended models of Llama-3-8B-Instruct and Mistral-7B-Instruct-v0.1 since these are the latest models for which SecAlign adapters were trained. We choose the equivalent defended models for StruQ for consistency.

\begin{table}[t]
    \centering
    
    \subfloat[Meta-Llama-3-8B-Instruct\label{tab:secalign_llama_3}]{
        \begin{minipage}[t]{0.48\textwidth}
            \centering
            \begin{tabular}{c|c|c}
            \Xhline{2\arrayrulewidth} % thicker than \hline
            \textbf{Budget (Config.)} & \textbf{GCG ASR} & \textbf{\att\  ASR} \\
            \Xhline{2\arrayrulewidth} % thicker than \hline
            20 (0, 20) & 12.5\% & \textbf{20\%}\\
            25 (5, 20) & 32.5\% & \textbf{57.5\%} \\
            30 (10, 20) & 35\% & \textbf{60\%} \\
            35 (15, 20) & 45\% & \textbf{72.5\%} \\
            \Xhline{2\arrayrulewidth} % thicker than \hline
            \end{tabular}
        \end{minipage}
    }
    \hfill
    \vspace{2pt}
    \subfloat[Mistral-7B-Instruct-v0.1\label{tab:secalign_mistral}]{
        \begin{minipage}[t]{0.48\textwidth}
            \centering
            \begin{tabular}{c|c|c}
            \Xhline{2\arrayrulewidth}
            \textbf{Budget (Config.)} & \textbf{GCG ASR} & \textbf{\att\ ASR} \\
            \Xhline{2\arrayrulewidth} % thicker than \hline
            20 (0, 20) & 32.5\% & \textbf{75\%} \\
            25 (5, 20) & 57.5\% & \textbf{77.5\%} \\
            30 (10, 20) & 60\% & \textbf{80\%} \\
            35 (15, 20) & 70\% & \textbf{82.5\%} \\
            \Xhline{2\arrayrulewidth} % thicker than \hline
            \end{tabular}
        \end{minipage}
    }
    
    \caption{Attack success rates (out of 40 examples) against the SecAlign models for a Strong Knowledge Adversary.}
    \label{tab:asrs_secalign}
\end{table}

\begin{figure}[t]
\centering
\begin{tikzpicture}
    \begin{axis}[
        width=0.48\textwidth,
        height=4cm,
        xlabel={Iterations},
        ylabel={Target Logprobs},
        legend pos=north east,
        legend style={font=\scriptsize, draw=none, fill=none},
        line width=1pt,
        grid style={dashed,gray!30},
    ]
    
    % Plot baseline from CSV
    \addplot[
        color=blue,
    ] table[x expr=\coordindex, y=baseline, col sep=comma] {data_raw/mistral_loss_curves.csv};
    
    % Plot attack from CSV
    \addplot[
        color=red,
    ] table[x expr=\coordindex, y=attack, col sep=comma] {data_raw/mistral_loss_curves.csv};
    
    \legend{Baseline (GCG), \att}
    \end{axis}
\end{tikzpicture}

\caption{The average loss curves (shown here for a budget of $30$ tokens for the SecAlign defended Mistral) show the ``warm start'' effect —- a good initializer from the first $350$ iterations significantly helps the usual GCG algorithm for the last $150$ iterations.}
\label{fig:combined_results}
\end{figure}

\noindent \textbf{Evaluation Dataset.} The security evaluation  of StruQ and SecAlign was originally performed on a test set of $208$ examples from the AlpacaFarm dataset which contain a non-empty ``\verb|input|" field. The attacker's prompt injection is placed at the end of the data section of the input. The prompt injection placed at the end itself consists of the payload, along with the optimized adversarial tokens. The attacker payload used by SecAlign and StruQ for their evaluations is the string ``\texttt{Print exactly 'Hacked!'}" with the target string `\verb|Hacked|'.

For our evaluation, we randomly chose $50$ out of the $208$ examples from the AlpacaFarm evaluation dataset. Due to hardware limitations, we only report our attack success rate out of $40$ examples that we could complete. These $40$ examples represent $\sim 20\%$ of the evaluation set.

\noindent \textbf{Evaluation Metrics.}
SecAlign and StruQ evaluated their defenses on a simple prompt injection task, where the attacker's goal is to cause the LLM to output the string ``\texttt{Hacked}".
The attack is considered successful if the output generated by the LLM starts with the string ``\texttt{Hacked}" or the string ``\texttt{hacked}". All generations happen at temperature $0$.

\noindent \textbf{Baseline.} We compare our attack with GCG as the baseline. In all our experiments, we control for all the following variables. 
\begin{enumerate}[itemsep=0pt, parsep=0pt, topsep=0pt, partopsep=0pt]
    \item Attacker budget and configuration (\textit{i.e.}, number of modifiable prefix and suffix tokens)
    \item Attack initialization
    \item Number of optimization iterations
    \item Number of forward passes on each iteration ($B$)
    \item Value of $p$ (\textit{i.e.} the number of top gradients chosen from)
\end{enumerate}
That is, we run both GCG and our attack for each of the test examples under the same conditions and budgets with the only difference being the optimization strategies.

\noindent \textbf{Implementational Parameters.} We run both attacks for $500$ iterations, with $B=512$ forward passes on each iteration. For both attacks, we take $p=256$ promising candidates for each token position. 
For the implementation of our loss function, we compute average sensitivities based on a subset of $50$ examples from the Dolly fine-tuning dataset \cite{DatabricksBlog2023DollyV2}, and drop the bottom $75\%$ of all attention heads.

\noindent \textbf{SecAlign - Attack Success Rate.} We perform our evaluation over a range of different attacker budgets and prefix-suffix configuratons. The results are as shown in \cref{tab:asrs_secalign}. With the default budget and configuration as in the original SecAlign paper, we can break $\mathbf{75\%}$ of the examples for the defended Mistral model and $\mathbf{20\%}$ of the examples for the defended Llama-3 model. Both of these numbers outperform the baseline success rate achieved by GCG. Further, by increasing the budget by just $5$ tokens, we can increase the success rate for the defended Llama-3 model to $\mathbf{57.5\%}$

\noindent \textbf{StruQ - Attack Success Rate.} StruQ represents a significantly weaker defense than SecAlign. Just by running GCG on the $40$ examples above with the default attacker budget of $20$ tokens in a suffix-only configuration, we were able to break $28$ examples for the StruQ defended Mistral and $23$ examples for the StruQ defended Llama-3. We apply \att\ to the remaining $12$ and $17$ examples with the same budget and configuration. \att\ breaks $\mathbf{10}$ out of the $12$ examples for the StruQ defended Mistral and $\mathbf{8}$ out of the $17$ examples for the StruQ defended Llama-3.

\subsection{Attack Success Rate - \att ++}

\noindent\textbf{Models.} - Due to resource limitations, we run the \att ++ experiments on only the two most robust defended models.
\begin{enumerate}[itemsep=0pt, parsep=0pt, topsep=0pt, partopsep=0pt]
    \item \textbf{SecAlign Meta-Llama-3-8B-Instruct} - We choose this model since this model shows the highest resistance to prompt injection attacks against Strong Knowledge Adversaries (\cref{tab:asrs_secalign}), thus providing the strongest challenge to a Weak Knowledge Adversary as well.
    \item \textbf{SecAlign++} - This is a production-grade Instruction Hierarchy model which represents a realistic target model for a Weak Knowledge Adversary whilst showing comparable robustness to the previous models, thus providing a hard target.
\end{enumerate}

\noindent\textbf{Training Dataset.} For each training run, we train our attacks on a training set of $10$ randomly chosen examples from the $208$ AlpacaFarm examples which have an \verb|input| field. Like before, we place the payload ``\verb|Print exactly 'Hacked!'|'' at the end of the input surrounded by a prefix and suffix.

\noindent\textbf{Evaluation Dataset.} We use the same dataset as for the Strong Knowledge Adversary case. That is, we use the remaining $198$ unseen examples from the AlpacaFarm dataset which contain an ``\verb|input|'' field as our test set. We use the same prompt format as expected by the SecAlign model. For the instruction-hierarchy model, we evaluate both modes of operation, one where we place the instruction at the highest trust level, that is, as the ``system'' prompt and place the data at the lowest trust level, that is, as the ``input'' and the other where we place the instruction as the ``user" and the data at ``input". This provides the worst-case scenario for our attacks.

\noindent\textbf{Evaluation Metrics.} For each training run for a fixed budget, (trained with $10$ examples), we compute the Attack Success Rate over the remaining $198$ examples by counting the number of model outputs that start with ``\verb|Hacked|'' or ``\verb|hacked|''.

\noindent For each attacker budget, we run $3$ different training runs, each with a completely different set of $10$ training examples to normalize for the effect of training set on the algorithm.

\noindent \textbf{Baseline.} We compare \att ++ with a universal version of GCG as the baseline. We note that a universal GCG over this combined loss function is equivalent to running an algorithm like NeuralExec ~\cite{pasquini2024neuralexeclearningand} with a simpler training objective.

\noindent \textbf{Implementational Parameters.}
For each attack run, like before, we control for initialization, number of iterations, number of forward passes per iteration, randomization strategy, and number of top gradients. In addition, we also control for the training dataset.
We run both algorithms for a total of $1000$ iterations with $p=256$ and $B=512$.
For \att++, we split the $1000$ iterations into $700$ steps of attention loss minimization and $300$ steps of Univ-GCG, with the sensitivities being refreshed every $50$ iterations (that is, $F=14, s_1=50, s_2=300$ in the notation of \cref{alg:astra_plus_plus}). We uniformly clip $50\%$ of the heads for all the loss functions.

\noindent \textbf{Results.} We show the mean and standard deviation for the testing attack success rate (averaged over 3 training and testing runs) of \att++ in \cref{tab:asrs_univ}.\ \att ++ significantly outperforms Universal GCG on average. Nonetheless, we observe that GCG can also create strong universal attacks, as can be seen by the high standard deviations which come due to outlier training runs which led to some success.

\begin{table}[t]
    \centering
    
    \subfloat[SecAlign defended Meta-Llama-3-8B-Instruct\label{tab:univ_secalign}]{
        \begin{minipage}[t]{0.48\textwidth}
            \centering
            \begin{tabular}{c|c|c}
            \Xhline{2\arrayrulewidth} % thicker than \hline
            \textbf{Budget (Config.)} &  
            \makecell{
            \textbf{Univ-GCG ASR}\\ Mean (Std. Dev.) } &
            \makecell{
            \textbf{\att++ ASR}\\ Mean (Std. Dev.) } \\
            \Xhline{2\arrayrulewidth} % thicker than \hline
            30 (15, 15) & 11\% (17\%) & \textbf{28\%} (48\%) \\
            35 (17, 18) & 55\% (48\%) & \textbf{96\%} (5\%) \\
            40 (20, 20) & 2\% (2\%) & \textbf{85\%} (18\%) \\
            \Xhline{2\arrayrulewidth} % thicker than \hline
            \end{tabular}
        \end{minipage}
    }
    \hfill
    \vspace{2pt}
    \subfloat[SecAlign++ defended model (instruction as ``system")\label{tab:univ_secalign_pp}]{
        \begin{minipage}[t]{0.48\textwidth}
            \centering
            \begin{tabular}{c|c|c}
            \Xhline{2\arrayrulewidth} % thicker than \hline
            \textbf{Budget (Config.)} &  
            \makecell{
            \textbf{Univ-GCG ASR}\\ Mean (Std. Dev.) } &
            \makecell{
            \textbf{\att++ ASR}\\ Mean (Std. Dev.) } \\
            \Xhline{2\arrayrulewidth} % thicker than \hline
            30 (15, 15) & 19\% (16\%) & \textbf{62\%} (53\%) \\
            35 (17, 18) & 22\% (21\%) & \textbf{87\%} (12\%) \\
            40 (20, 20) & 30\% (33\%) & \textbf{55\%} (20\%) \\
            \Xhline{2\arrayrulewidth} % thicker than \hline
            \end{tabular}
        \end{minipage}
    }
    \hfill
    \vspace{2pt}
    \subfloat[SecAlign++ defended model (instruction as ``user")\label{tab:univ_secalign_pp2}]{
        \begin{minipage}[t]{0.48\textwidth}
            \centering
            \begin{tabular}{c|c|c}
            \Xhline{2\arrayrulewidth} % thicker than \hline
            \textbf{Budget (Config.)} &  
            \makecell{
            \textbf{Univ-GCG ASR}\\ Mean (Std. Dev.) } &
            \makecell{
            \textbf{\att++ ASR}\\ Mean (Std. Dev.) } \\
            \Xhline{2\arrayrulewidth} % thicker than \hline
            30 (15, 15) & 1\% (11\%) & \textbf{69\%} (1\%) \\
            35 (17, 18) & \textbf{31\%} (24\%) & 22\% (54\%) \\
            40 (20, 20) & 22\% (53\%) & \textbf{37\%} (30\%) \\
            \Xhline{2\arrayrulewidth} % thicker than \hline
            \end{tabular}
        \end{minipage}
    }

    \caption{Test ASR on 198 unseen examples (mean and standard deviations over 3 training runs on distinct training datasets)}
    \label{tab:asrs_univ}
\end{table}

\begin{figure}[t]
\centering
\begin{center}
\footnotesize
    \begin{tabular}{@{}c@{\hspace{0.3em}}l@{\hspace{2em}}c@{\hspace{0.3em}}l@{\hspace{2em}}c@{\hspace{0.3em}}l@{}}
\textcolor{yellow}{\rule{0.8cm}{1.5pt}} & Only first &
\textcolor{orange}{\rule{0.8cm}{1.5pt}} & Only last &
\textcolor{red}{\rule{0.8cm}{1.5pt}} & Uniform \\
\textcolor{green}{\rule{0.8cm}{1.5pt}} & Avg. Sen. &
\textcolor{black}{\rule{0.8cm}{1.5pt}} & Clip. Sen. &
\textcolor{blue}{\rule{0.8cm}{1.5pt}} & Baseline (GCG) \\
\end{tabular}
\end{center}
    % \begin{minipage}{0.48\textwidth}
    %     \centering
    %     \begin{tikzpicture}
    %         \begin{axis}[
    %             width=0.9\linewidth,
    %             height=5cm,
    %             xlabel={Iterations},
    %             ylabel={Target Logprobs},
    %             legend pos=south west,
    %             legend style={font=\scriptsize, draw=none, fill=none},
    %             line width=1pt,
    %             ymin=0,
    %             ymax=35
    %         ]
            
    %         % Plot baseline from CSV
    %             \addplot[
    %                 color=yellow,
    %             ] table[x expr=\coordindex, y=only_first, col sep=comma] {data_raw/ablations_meta.csv};
    
    %             \addplot[
    %                 color=orange,
    %             ] table[x expr=\coordindex, y=only_last, col sep=comma] {data_raw/ablations_meta.csv};
    
    %             \addplot[
    %                 color=red,
    %             ] table[x expr=\coordindex, y=uniform, col sep=comma] {data_raw/ablations_meta.csv};
    
    %             \addplot[
    %                 color=green,
    %             ] table[x expr=\coordindex, y=avg_sens, col sep=comma] {data_raw/ablations_meta.csv};
    %             \addplot[
    %                 color=black,
    %             ] table[x expr=\coordindex, y=clipped_sens, col sep=comma] {data_raw/ablations_meta.csv};

    %             \addplot[
    %                 color=blue,
    %             ] table[x expr=\coordindex, y=baseline, col sep=comma] {data_raw/ablations_meta.csv};
                
    %             % \legend{Only first, Uniform, Avg. Sen., Clip. Sen., Baseline (GCG)}
            
    %         \end{axis}
    %     \end{tikzpicture}
    % \end{minipage}

    \begin{minipage}{0.48\textwidth}
        \centering
        \begin{tikzpicture}
            \begin{axis}[
                width=0.9\linewidth,
                height=4cm,
                xlabel={Iterations},
                ylabel={Target Logprobs},
                legend pos=south west,
                legend style={font=\scriptsize, draw=none, fill=none},
                line width=1pt,
                ymin=0,
                ymax=35
            ]
            
            % Plot baseline from CSV
                \addplot[
                    color=yellow,
                ] table[x expr=\coordindex, y=only_first, col sep=comma] {data_raw/ablations_mistral.csv};
    
                \addplot[
                    color=orange,
                ] table[x expr=\coordindex, y=only_last, col sep=comma] {data_raw/ablations_meta.csv};
    
                \addplot[
                    color=red,
                ] table[x expr=\coordindex, y=uniform, col sep=comma] {data_raw/ablations_mistral.csv};
    
                \addplot[
                    color=green,
                ] table[x expr=\coordindex, y=avg_sens, col sep=comma] {data_raw/ablations_mistral.csv};
                \addplot[
                    color=black,
                ] table[x expr=\coordindex, y=clipped_sens, col sep=comma] {data_raw/ablations_mistral.csv};
    
                \addplot[
                    color=blue,
                ] table[x expr=\coordindex, y=baseline, col sep=comma] {data_raw/ablations_mistral.csv};
                
                % \legend{Only first, Uniform, Avg. Sen., Clip. Sen., Baseline (GCG)}
            
            \end{axis}
        \end{tikzpicture}
    \end{minipage}

\caption{Clipped sensitivities outperform other weighting functions - shown here for the SecAlign defended Mistral}
\label{fig:mist_ablations}
\end{figure}

\subsection{Ablation on Weights}

The key novelty in the construction of \att\ lies in the loss function. We perform an ablation study to show that the clipped sensitivities used as weights in our loss function outperform other intuitive and obvious choices for weights such as uniform weights and weighing only the first and last layer in addition to outperforming the baseline itself. 

Concretely, we take the $40$ examples from the evaluation dataset of \att\ and run \att\ on them with $5$ different weighting functions along with the baseline. Like before, we control for all the confounders like initialization, attacker budget, and other attack hyperparameters.
The average loss curves for the SecAlign defended version of Mistral in \cref{fig:mist_ablations} show that clipped sensitivities outperform other weighting functions.

\section{Discussion}
\label{sec:discussion}

\subsection{A more general framework}

While \att\ maximizes the attention on the payload, it can be generalized to distances between attention matrices.

Our attention loss defined in \cref{eq:combined_att_loss} can be written as 
$$\mathrm{AttLoss}_i^{(l)}(\mathbf{x}, y) = \sum_{j \in J} \frac{1}{|J|} - A_i^{(l)}(\mathbf{x})[n][j]$$

This can be interpreted as measuring the difference between an ``ideal'' attention pattern (which treats all tokens in the payload equally) and the ``true'' attention pattern. Since rows of attention matrices have multiple interpretations as probability masses and convex combinations (\cref{sec:attack}), instead of looking at the difference between attention values, we could use more sophisticated distances such as Wasserstein distances to craft attacks. We leave the exploration of such attacks to future work.

\subsection{Limitations of \att}

\noindent \textbf{Performance.} The primary limitation of \att\ is its memory footprint and performance. \att\ relies on computing on attention matrices during a model's forward and backward passes, which is a memory intensive operation. As a result, there's significant slowdown in the entire attack. We found \att\ to be roughly $1.7-2.5\times$ slower than GCG. We leave the question of time efficiency to future work.

\noindent \textbf{Dynamic Prompt Injections.} We evaluated \att\ and \att ++ for the simple case where the target string is a fixed string. In general, prompt injections can require dynamically generated target strings such as when the attacker wants to leak private data based on the conversation history. We leave the extension of our attack to dynamic target strings for future work.

\subsection{Evaluation of Prompt Injection Defenses}

Our results (\cref{sec:eval}) show that even crude attacks can break prompt injection defenses with increased budgets. We propose that prompt injection defenses should, instead of evaluating at one single budget, show their robustness under scaling as that better captures the realities of the threat model.

% \noindent \textbf{Failure mode: Degraded instruction following capability} Our attack works on the assumption that a model can follow an attacker's instruction when there's no surrounding context and all the attention is concentrated on the attacker's payload. In practice, with the SecAlign models, we find that this assumption isn't entirely valid. To assert this, we perform two simple experiments where we give three simple and explicit instructions $\verb|Print exactly `Hacked!'|$, $\verb|Output |$ in the instruction field of an input and compare the probabilities of the string $\verb|`Hacked'|$ appearing in the output.
\section{Related Work}
\label{sec:related_work}

\noindent \textbf{Jailbreaks.} Jailbreaks are attacks where attackers try to cause the model to output content forbidden by its content safety policy. While several approaches exist across white-box, black-box and grey-box threat models ~\cite{mehrotra2024treeattacksjailbreakingblackbox,sitawarin2024palproxyguidedblackboxattack,hayase2024querybasedadversarialpromptgeneration}, a major class of algorithms optimize the input to force the model to begin its answer with a target string such as ``Sure". In the whitebox setting, the SOTA solution to this problem is achieved by GCG and its variants and enhancements, such as Momentum-GCG \cite{zhang2025boostingjailbreakattackmomentum}, AutoDAN \cite{zhu2023autodaninterpretablegradientbasedadversarial}, and Accelerated GCG \cite{zhao2024acceleratinggreedycoordinategradient}.

\noindent \textbf{Optimization-based Prompt Injections.} Since the optimization problem to be solved for jailbreaking is similar to a prompt injection, GCG has also been adapted for generating prompt injections in the whitebox setting. NeuralExec \cite{pasquini2024neuralexeclearningand} and Universal \cite{liu2024automaticuniversalpromptinjection} use versions of GCG to generate prompt injections. Imprompter ~\cite{fu2024impromptertrickingllmagents} used GCG to leak private data.

Crucially, they all fundamentally rely on gradients of target logprobs to guide their optimizations. In contrast, \att\ uses the gradients in an architecture-aware way, by incorporating them into the weights for attention heads. In addition, our approach is specifically tailored for prompt injections, which allows us to achieve high success rates even in a realistic threat model.

\noindent \textbf{Attacks manipulating Attention.}
Attn-GCG~\cite{wang2024attngcgenhancingjailbreakingattacks} uses attention matrices to craft jailbreaks. Our work focuses on attacks against fine-tuning-based prompt injection defenses. There are several key differences between \att\ and Attn-GCG. First, their loss function adds a regularization term to the standard GCG loss. In contrast, our algorithm uses attention as a warm start for GCG. Second, they focus attention on the gibberish adversarial suffix, whereas, \att~focuses attention on human-readable prompt injection instructions. Third, Attn-GCG only considers the last decoder layer without appropriately chosen weights. As we show in \cref{fig:mist_ablations}, this doesn't work well. Thus, Attn-GCG shows only marginal improvements. By contrast, \att~shows much higher performance on injection tasks. 

In the vision domain, AICAttack~\cite{MIR-2024-09-397} uses attention to fool captioning algorithms. Guo et al. also use attention to craft backdoor attacks in vision transformers and BERT-based encoder models~\cite{guo2024backdoorattackvisiontransformers,lyu-etal-2023-attention}. We leave attacking multi-modal language models to future work, though we expect our techniques to transfer. Our work focuses on text LLMs that are the most widely-deployed form of language models today.

\section{Conclusion}
\label{sec:conclusion}

Prompt injection attacks are a major risk against LLM-integrated systems. Fine-tuning based defenses teach LLMs to distinguish instructions and data. We provided an adaptive security analysis of this class of defenses by analyzing three recent representative systems (StruQ, SecAlign and SecAlign++). While these systems included an adaptive security analysis, they were limited by weak attacks based on Greedy Co-ordinate Gradient. We showed that such attacks make ineffective use of whitebox access to the model. We created the \att~attack that uses architectural information about the model. \att~shows high success rates against these defenses, both in the original evaluation setting and an extended evaluation setting that involves universality across prompts. We believe that our work has set a foundation for the correct evaluation of fine-tuning based prompt injection defenses.

\bibliography{references} % Name of your .bib file
\appendix
\section{Full Algorithm - \att ++}
\label{app:astra_plus_plus}

\begin{algorithm}[!h!]
    \SetAlgoLined
    \KwIn{Training dataset of contexts $\mathcal{D} = \{\mathbf{x}_{\mathrm{Trusted}} \Vert \mathbf{x}_{\mathrm{Payload}}\}$}
    \KwIn{Initial adversarial perturbation $\mathbf{r}^{(0)} \in V^{|\mathbf{r}^{(0)}|}$}
    \KwIn{Target sequence of tokens $\mathbf{y} = (y_1, y_2, \dots, y_m) \in V^{m}$}
    \KwIn{A function Loss: $V^{*} \times V^{m} \to \mathbb{R}$}
    \textbf{Parameters: } $p \in \mathbb{N}$ — number of top candidates to consider; $B \in \mathbb{N}$ — number of evaluations per step; $N \in \mathbb{N}$ — number of optimization steps\\
    \KwOut{List of Perturbations obtained during optimization and their corresponding Average Logprobs}

    \Begin{
        AvgLogprobs $\leftarrow [\ ]$ \;
        Perturbations $\leftarrow [\ ]$ \;

        \For{$k \leftarrow 0$ \KwTo $N-1$}{
            \ForEach{$i \in |\mathbf{r}^{(0)}|$}{
                $R_i^{(k)} \leftarrow \mathrm{Top\text{-}p} \left( - \sum \limits_{\mathbf{x}\in\mathcal{D}} \frac{\nabla_{e_{x_i}} \mathrm{Loss}(\mathbf{x} \Vert \mathbf{r}^{(k)}, \mathbf{y})}{\lVert \nabla_{e_{x_i}} \mathrm{Loss}(\mathbf{x} \Vert \mathbf{r}^{(k)}, \mathbf{y}) \rVert} \right)$
            }

            \For{$b \leftarrow 1$ \KwTo $B$}{
                $i \sim \mathrm{Uniform}(\{1, 2, \dots, |\mathbf{r}|\})$ \\
                $\tilde{\mathbf{r}}^{(k)}_{b} \leftarrow \mathbf{r}^{(k)}$ \\
                $\tilde{\mathbf{r}}^{(k)}_b[i] \sim \mathrm{Uniform}(R_i^{(k)})$
            }

            $b^* \leftarrow \arg\min_b\ \sum\limits_{\mathbf{x}\ \in \mathcal{D}} \mathrm{Loss}(\mathbf{x}\Vert\tilde{\mathbf{r}}^{(k)}_{b},\mathbf{y})$ \\
            $\mathbf{r}^{(k+1)} \leftarrow \tilde{\mathbf{r}}^{(k)}_{b^{*}}$\\
            AvgLogprobs$[k] \leftarrow \overline{\mathrm{Logprobs}}_{\mathcal{D}, \mathbf{y}}(\mathbf{r}^{(k+1)})$\\
            Perturbations$[k] \leftarrow \mathbf{r}^{(k+1)}$
        }

        \Return Perturbations, AvgLogprobs
    }
    \caption{$\texttt{\textbf{UnivGenGCG}}_{p, B, N}(\mathcal{D}, \mathbf{r}^{(0)} , \mathbf{y}, \mathrm{Loss})$}
    \label{alg:univ_gcg}
\end{algorithm}

\begin{algorithm}[!h!]
    \SetAlgoLined

    \KwIn{Training dataset of contexts $\mathcal{D} = \{\mathbf{x}_{\mathrm{Trusted}} \,\Vert\, \mathbf{x}_{\mathrm{Payload}}\}$}
    \KwIn{Initial perturbation $\mathbf{r}^{(0)}$}
    \KwIn{Target tokens $\mathbf{y} = (y_1, y_2, \dots, y_m) \in V^{m}$}
    \textbf{Parameters: } 
    $p \in \mathbb{N}$ — number of top candidates to consider; 
    $B \in \mathbb{N}$ — number of evaluations per step; 
    $F \in \mathbb{N}$ — number of local sensitivity recomputations; 
    $s_1$ — number of steps optimizing each local attention loss; 
    $s_2$ — number of steps optimizing the final target logprobs. \\
    \KwOut{Optimized perturbation $\mathbf{r}^{*}$}

    \Begin{
        $\mathrm{AllAvgLogprobs} \leftarrow [\ ]$ \;
        $\mathrm{AllPerturbations} \leftarrow [\ ]$ \;

        \For{$c \in \{1, 2, \dots, F\}$}{
            $\mathrm{Perturbations}_c,\ \mathrm{AvgLogprobs}_c 
            \leftarrow \texttt{\textbf{UnivGenGCG}}_{p, B, s_1}\big(\mathcal{D}, \mathbf{r}^{*}_{c-1}, \mathbf{y}, \overline{\mathrm{AttLoss}}_{\mathbf{r}_c^{*}, \mathcal{D}}\big)$

            $\mathbf{r}_{c}^{*} \leftarrow \mathrm{Perturbations}_c[-1]$ \;
            $\mathrm{AllAvgLogprobs} \mathrel{+}= \mathrm{AvgLogprobs}_c$ \;
            $\mathrm{AllPerturbations} \mathrel{+}= \mathrm{Perturbations}_c$ \;
        }

        $\mathbf{r}^* \leftarrow \mathrm{AllPerturbations}[\arg\min\ \mathrm{AllAvgLogprobs}]$ \;

        $\mathrm{Perturbations},\ \mathrm{AvgLogprobs} 
        \leftarrow \texttt{\textbf{UnivGenGCG}}_{p, B, s_2}(\mathcal{D}, \mathbf{r}^{*}, \mathbf{y}, \mathrm{TargetLogprobs})$ \;

        $\mathbf{r}^{*} \leftarrow \mathrm{Perturbations}[\arg\min\ \mathrm{AvgLogprobs}]$ \;
        \Return $\mathbf{r}^{*}$
    }

    \caption{$\texttt{\textbf{\att ++}}_{p, B, F, s_1, s_2}(\mathcal{D}, \mathbf{r}^{(0)}, \mathbf{y})$}
    \label{alg:astra_plus_plus}
\end{algorithm}

\end{document}